\documentclass[CRPHYS,Unicode,manuscript]{cedram}

\title{Generalizations of Parisi's replica symmetry breaking and overlaps in random energy models}

\alttitle{G\'{e}n\'{e}ralisations de la brisure de sym\'{e}trie des r\'{e}pliques de Parisi et des overlaps dans les mod\`{e}les d’\'{e}nergies al\'{e}atoires}

\author{\firstname{Bernard} \lastname{Derrida}\IsCorresp \CDRorcid{0000-0001-6994-0226}}
\address{Coll\`ege de France, 11 place Marcelin Berthelot,  75005 Paris, France}
\address{Laboratoire de Physique de l'Ecole Normale Sup\'erieure, ENS, Universit\'e PSL, CNRS, Sorbonne Universit\'e,  Universit\'e de Paris, F-75005 Paris, France}
\email[B. Derrida]{bernard.derrida@college-de-france.fr}

\author{\firstname{Peter} \lastname{Mottishaw}\CDRorcid{0000-0002-0091-4094}}
\address{SUPA, School of Physics and Astronomy, University of Edinburgh, Peter Guthrie Tait Road, Edinburgh EH9 3FD, United Kingdom}
\email[P. Mottishaw]{peter.mottishaw@ed.ac.uk}

\keywords{Disordered systems, Spin glasses, Replica symmetry breaking, Random Energy Model}

\begin{abstract}
  The random energy model (REM) is the simplest spin glass model which exhibits replica symmetry breaking. It is well known since the 80's that its overlaps are non-selfaveraging and that their  statistics   satisfy the predictions of the replica theory.  All these statistical properties  can be understood by considering that the low energy levels are the points generated by a Poisson process with an exponential density. Here we first show how,  by replacing the exponential density by a sum of two exponentials, the overlaps statistics are modified.    One way to reconcile these results with the replica theory is to allow the blocks in the Parisi matrix to fluctuate. Other examples where the sizes of these blocks should fluctuate include the finite size corrections of the REM, the case of discrete energies and the overlaps between two temperatures. In all these cases, the blocks sizes not only  fluctuate but need to take complex values if one wishes to reproduce the results of our replica-free calculations.
\end{abstract}

\begin{altabstract}
    Le mod\`{e}le d'\'{e}nergies al\'{e}atoires (REM) est le mod\`{e}le de verre de spin le plus simple qui pr\'{e}sente une brisure de sym\'{e}trie des r\'{e}pliques. Il est bien connu depuis les ann\'{e}es 80 que ses overlaps ne sont pas automoyennants et que leurs statistiques sont celles prédites par la méthode des r\'{e}pliques. Ces propri\'{e}t\'{e}s statistiques peuvent \^{e}tre comprises en consid\'{e}rant que les niveaux d'\'{e}nergie les plus bas sont les points g\'{e}n\'{e}r\'{e}s par un processus de Poisson de densit\'{e} exponentielle. Nous montrons ici dans un premier temps comment ces statistiques d'overlaps sont modifi\'{e}es lorsqu'on remplace la densit\'{e} exponentielle par une somme de deux exponentielles. Une fa\c{c}on de concilier ces r\'{e}sultats avec la th\'{e}orie des r\'{e}pliques est de permettre aux blocs de la matrice de Parisi de fluctuer. D'autres exemples o\`{u} la taille de ces blocs doit fluctuer incluent les corrections de taille finie du REM, le cas des \'{e}nergies discr\`{e}tes et les overlaps entre deux temp\'{e}ratures. Dans tous ces cas, non seulement la taille des blocs fluctue mais elle doit prendre des valeurs complexes si l'on souhaite reproduire les r\'{e}sultats de nos résultats obtenus directement, c'est à dire sans utiliser la méthode des répliques. 
\end{altabstract}

\begin{document}
\maketitle

\section{Introduction}
The introduction 
by Edwards and Anderson \cite{Edwards_1975_Theory} in 1975 
of the replica method 
and of  the overlaps as an order parameter 
was a big step in the theory of spin glasses \cite{Parisi_2004_OVERLAP}. 
For a spin glass  model  with $N$ Ising spins, the overlap  $q ({\bf S,S'}) $  between two spin configurations 
${\bf S} \equiv \{S_i \pm 1\}$ and
${\bf S'} \equiv \{S_i' \pm 1\}$
 is defined as
\begin{equation}
q({\bf S,S'} )  ={1 \over N} \sum_{i=1,N}  S_i \, S_i'
\label{A1}
\end{equation}
Qualitatively, the  idea  was that, because spin glasses have a rugged energy landscape,   two  typical spin configurations  at equilibrium have, at low temperature,  the tendency  of being trapped in the same valley giving rise to a non-zero overlap. Quantitatively,
for  a spin glass model with  quenched pair
   interactions ${\bf J}\equiv \{J_{i,j}\}$  
\begin{equation}
E_{\bf J }( 
{\bf S}) = - \sum_{i,j} J_{i,j} S_i S_j
\label{A2}
\end{equation}
  the overlaps  can be characterized by their probability  distribution
\begin{equation}
P_{\bf J }(Q) =  { \sum_{\bf S} \sum_{\bf S'} \delta\Big(q({\bf S},{\bf S'})-Q \Big) 
\ \exp\left[-\beta  \Big(E_{\bf J}( {\bf S})+ E_{\bf J}( {\bf S}) \Big) \right] \over
 \sum_{\bf S} \sum_{\bf S'}  
\ \exp\left[-\beta  \Big(E_{\bf J}( {\bf S})+ E_{\bf J}( {\bf S}) \Big) \right] }
\label{A3}
\end{equation}
where $\beta$ is the inverse temperature.
As long as the system is finite ($N < \infty$) this distribution $P_{\bf J }(Q)$ is  a broad  function of $Q$ and depends on the sample ${\bf J}$. However in the thermodynamic limit, it was initially expected that, in the spin-glass phase and for almost all samples ${\bf J}$,  it becomes a sum of two delta functions
\begin{equation}
\lim_{N \to \infty} P_{\bf J }(Q) =   {1 \over 2} \left[ \delta(Q-q_{EA}) + \delta(Q+q_{EA}) \right]
\label{A4}
\end{equation}
and in \cite{Edwards_1975_Theory}  a mean field theory was developed to determine  the value of the Edwards-Anderson order parameter $q_{EA}$.
(if some odd interactions were added to (\ref{A2}) the limit (\ref{A4}) would reduce to a single delta function 
$\lim_{N \to \infty} P_{\bf J }(Q) =    \delta(Q-q_{EA}) $).

Soon after the   Edwards Anderson 1975 paper \cite{Edwards_1975_Theory}, Sherrington and Kirkpatrick \cite{Sherrington_1975_Solvable} considered the infinite range version of the model (\ref{A2}), for which the mean field approximation was expected to become exact. Using the replica approach, they could obtain explicit expressions of
$q_{EA}$ and of
 the average free energy. However they realized that their analytic solution could not be correct because,  at low temperature,   it leads to a negative entropy  \cite{Sherrington_1975_Solvable} and to a negative variance of the free-energy \cite{Toulouse_1981_Free}

It was only in 1979 that Parisi was able to  overcome these difficulties with a Replica Symmetry Breaking (RSB) scheme which turned out to  give  the correct free energy for the Sherrington Kirkpatrick model.
Initially, the solution was based on unconventional mathematics such as using matrices of non-integer size or  replacing maxima by minima.
It took then a few more years before the physical interpretation of Parisi's solution was understood \cite{Parisi_1983_Order,Mezard_1984_Nature,Mezard_1984_Replica,Mezard_1987_Spin} and even longer before it was confirmed  by a series of rigorous  mathematical proofs \cite{Guerra_2003_Broken,Talagrand_2006_Parisi,Arguin_2009_structure,Panchenko_2013_Parisi}. See \cite{Charbonneau_2023_Spin} for a recent review of the theory of spin glasses and replica symmetry breaking.
Besides an analytic way  of calculating the exact free energy of the Sherrington-Kirkpatrick model, Parisi's solution led to a number of surprising predictions.  One of them was that, instead of (\ref{A4}),  the distribution $P_{\bf J }(Q)$ remains broad  and sample dependent even in the thermodynamic limit ($N \to \infty$). Moreover it gave an explicit way of calculating these fluctuations
	\cite{Mezard_1984_Nature,Mezard_1984_Replica,Mezard_1987_Spin}.

\begin{figure}[h]
 \centerline{\includegraphics[width=10.5cm]{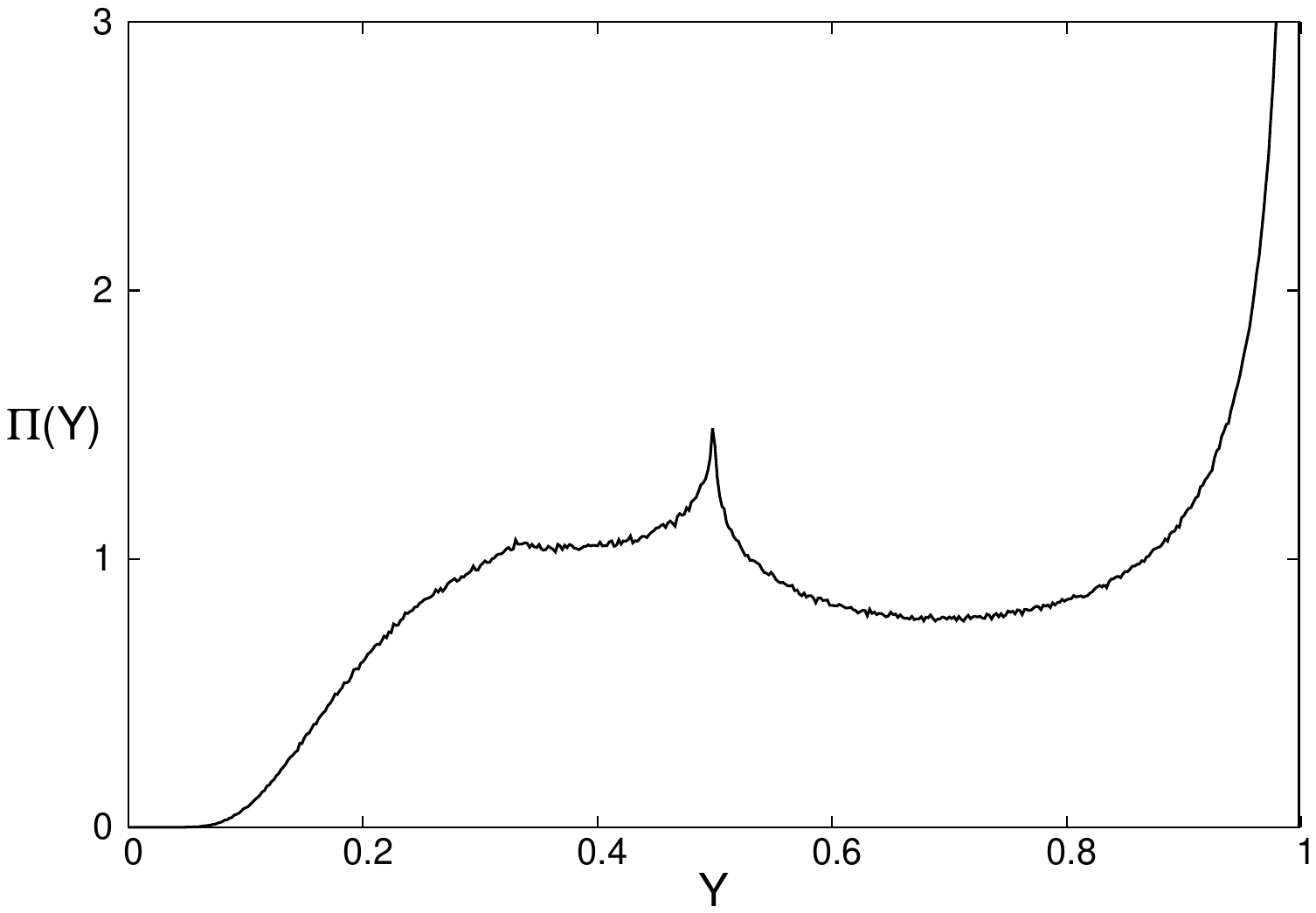}}
 \caption{The distribution $\Pi_\mu (Y)$ as a function of $Y$ in the case $\mu=1/3$}
 \label{figure1}
 \end{figure}

One way of describing  these fluctuations is to consider the   cumulative function
\begin{equation}
Y(Q)= \lim_{N \to \infty}\int_Q^1  P_{\bf J }(Q')  dQ'
\label{A5}
\end{equation}
which represents the probability that, at equilibrium, two configurations in the same energy landscape have an overlap  larger than $Q$. According to the RSB theory (see Appendix A),  $Y(Q)$ remains sample dependent  even in the thermodynamic limit and  its probability distribution $\Pi_\mu(Y)$ is universal for all the models which can be solved by the RSB. This distribution $\Pi_\mu(Y)$ is indexed by a single parameter  $0\leq \mu \leq 1 $ which depends  on $Q$,  on the temperature, on the magnetic field and on all the other parameters which may characterize a specific model. As a consequence if $ \langle Y \rangle $  is known for a given system (where $\langle . \rangle$ denotes the average over the samples, i.e. over  the ${\bf J}$'s)   all the moments of $Y$ are known. 
For example if  $ \langle Y \rangle=1-\mu $, one has 
\begin{equation}
\label{A6}
\langle Y \rangle =  1-\mu\ \ \ \ \ ; \ \ \ \ \
\langle Y^2 \rangle = {(1-\mu) (3-2 \mu)\over 3}  \ \ \ \ \ ; \ \ \ \ \
\langle Y^3 \rangle = {(1-\mu)(15-17 \mu + 5 \mu^2 )\over 15}
\end{equation}
A typical shape 
of the distribution   $\Pi_\mu(Y)$ 
is shown in Figure \ref{figure1} 
 \cite{Mezard_1984_Replica,Derrida_1987_Statistical}
\\ \ \\ 
More generally one can consider,  for a given sample $\bf J$,  the probability $Y_k(Q)$ that  $k$ configurations have all their overlaps larger than $Q$, or the probability $Y_{k,k'}(Q)$  that, for two  groups  of $k$ and $k'$  configurations,  all the overlaps between pairs of configurations inside each group are larger than $Q$ but all pairs between two groups have an overlap less than $Q$.   According to the RSB theory (see Appendix A), the averages of these $Y_k$  or $Y_{k,k'} $ over the samples have also explicit expressions in terms of the parameter $\mu$
\begin{equation}
\label{A7}
\langle Y_k \rangle = 
{\Gamma(k-\mu) \over \Gamma(k) \ \Gamma(1-\mu) } 
 \ \ \ \ \ ; \ \ \ \ \ \langle Y_{k,k'} \rangle =
{\mu \ \Gamma(k-\mu) \ \Gamma(k'-\mu)  \over \Gamma(k+k') \ \Gamma(1-\mu)^2 } 
\end{equation}
These satisfy particular cases of the
Ghirlanda-Guerra relations
\cite{Ghirlanda_1998_General}: for example it is easy to check that
$$ 
\langle Y_{k+1} \rangle
 = 
{\langle Y_{2} \rangle \, 
\langle Y_{k} \rangle +
(k-1)\langle Y_{k} \rangle 
\over k}
$$

All these statistical properties   of the $Y_k$'s were first derived in 1984 using the replica method (see Appendix A). 
In a joint work 
\cite{Derrida_1985_Sample}
with G\'erard Toulouse in 1985,  they were confirmed by a replica-free calculation for the Random Energy Model (REM).

The goal of the present paper is to exhibit several simple models where these statistical properties   are no longer valid and have to be modified. As explained in section 3, these models are cousins of the Random Energy Model, in the sense that the energies of the configurations are independent random variables and that their properties can be obtained using replica-free methods.  Based on some exact expressions (\ref{D8}-\ref{D9}) and (\ref{D11}-\ref{D12}) which replace (\ref{A7}) for these models, we will discuss possible generalizations of the Parisi matrix (see the appendix A) where the sizes of the blocks have to fluctuate (sometimes with complex sizes !).

The paper is organized as follows. In section 2, we recall some well known properties of the REM
and why, in  the low temperature phase, the energies can be generated by a Poisson process with an exponential density.
In section 3, we establish some general expressions allowing us to calculate non-integer moments of the partition function as well as average overlaps  $\langle Y_k\rangle, \langle Y_{k,k'}  \rangle $ for a Poisson  process with an arbitrary density. In the case of an exponential density, this allows one to recover (\ref{A7}).
In section 4 we consider the case where the density  of the Poisson process is the sum of two exponentials and show  how  (\ref{A7}) are modified.  We will then suggest a way of  adapting the RSB approach   to reproduce these new expressions.
This is probably the simplest case for which the size of the blocks in  the Parisi matrix have  to fluctuate.
A straightforward generalization to the case of a sum of an arbitrary number of exponentials will allow us  to recover, in a unified way, earlier results  on finite size corrections and in the case of discrete energies.
In section 5 we generalize the approach to consider the  overlaps between two configurations at different  temperatures, and show that, again,  the size of the blocks of the Parisi matrix have to fluctuate. 
Finally In section 6, we discuss the replica method 
for a Poisson REM with
an arbitrary density of energy, and see how the replica scheme has to be modified in this case.

 \section{Short review of known results on the REM}
In this section, we first recall a few known features of the REM, in particular the fact that, in the thermodynamic limit,     the low temperature properties can be understood by considering that the energies are the realization of a Poisson process with an exponential density (\ref{B15}).
\subsection{The REM with Gaussian energies} 
The Random Energy model  was introduced \cite{Derrida_1980_Random,Derrida_1981_Random}   in 1980  as a simple  spin glass model which  exhibits a spin glass transition and which can be solved without  resorting to the replica trick. As for other Ising spin glass models like the Edwards Anderson or the Sherrington-Kirkpatrick models,  there are $2^N $ spin configurations  ${\bf S}$ whose energies are  Gaussian  random variables 
\begin{equation}
P(E) = \frac{1 }{ \sqrt{N   \pi} \, J}   \exp\left[ -\frac{ E^2 }{ N \,  J^2}  \right]
\label{B1}
\end{equation}
We will set $J=1$, for simplicity, in the rest of the paper.
The only difference with other spin glass models is that, in the REM, the energies of different configurations are independent random variables. Therefore, instead of the interactions ${\bf J}$ in (\ref{A2}),  a given sample ${\bf J}$ in the REM is specified  by $2^N$ random quenched variables $E({\bf S})$ distributed according to (\ref{B1}).
Then the partition function $Z(\beta)$ becomes a sum of $2^N$ independent random variables making many calculations much easier than for other spin-glass models
\begin{equation}
Z(\beta)=\sum_{{\bf S}=1}^{2^N} e^{- \beta E({\bf S})}
\label{B2}
\end{equation}
Still the REM exhibits a phase transition \cite{Derrida_1980_Random,Derrida_1981_Random} at some $\beta_c$.  
\begin{equation}
\beta_c = 2 \sqrt{\log2}
\label{B3}
\end{equation}
and, in the thermodynamic limit,  the average free is given by
\begin{equation}
\lim_{N \to \infty}  \frac{\langle \log Z(\beta)\rangle  }{ N}  =
\left\{ \begin{array}{lll} \log 2 + \frac{\beta^2 J^2 }{ 4} & \text{ for} &  \beta< \beta_c
\\  & & \\
\beta \,  \sqrt{\log2}  & \text{for} & \beta > \beta_c
\end{array} \right.
\label{B3a}
\end{equation}
where $\langle.\rangle$ denotes an average over the energies $E({\bf S})$. These properties have been confirmed in a number of rigorous works \cite{Eisele_1983_third,Galves_1989_Fluctuations,Olivieri_1984_existence,Bovier_2002_Fluctuations}; for a review see \cite{Bolthausen_2007_Random} or \cite{Kistler_2015_Derridas}. Note that the REM is an example of sums of exponentials of random i.i.d. variables which have been considered in several other contexts \cite{Pastur_1989_limit,BenArous_2005_Limit}. 

Many other properties of the REM have been calculated (see \cite{Derrida_2023_Random} for a review) like, finite size  corrections \cite{Derrida_1981_Random,Campellone_1995_Some,Campellone_2009_Replica,Derrida_2015_Finite},  the effect of a magnetic field \cite{Derrida_1981_Random}, the  location of the zeroes in the complex  plane of $\beta$ \cite{Moukarzel_1991_Numerical,Derrida_1991_zeroes,Saakian_2000_Random}, Fisher zeroes and the fluctuations of the spectral form factor of chaotic systems \cite{Bunin_2023_Fisher}, the effect of discrete energies \cite{Ogure_2005_exact,Jana_2007_Contributions,Derrida_2022_discrete}  or the integer and the non-integer moments of the partition function \cite{Gardner_1989_probability}. 

For example,  in the low temperature phase  (i.e. $\beta > \beta_c$), the  negative  moments  are given \cite{Gardner_1989_probability}   for large $N$ by 
\begin{equation}
\left\langle Z(\beta)^n \right\rangle = \left( {A \over \beta_c} \  \Gamma(1-\mu)  \right)^{n \over \mu} {\Gamma\left(1-{n \over \mu}\right) \over \Gamma(1-n)}   \ \ \ \ \text{ for} \ \ \ \ n <0 
\label{B4}
\end{equation}
where 
\begin{equation}
A =  {e^{N \beta_c \sqrt{\log2} } \over \sqrt{ \pi N} }
\label{B5}
\end{equation}
 and 
\begin{equation}
\mu={\beta_c \over \beta} 
\label{B6}
\end{equation}

In the low temperature phase (see (\ref{B3a})) the extensive part of the energy is constant. This is because the only configurations ${\bf S}$ which really contribute to the free energy are those  whose energies are very close to the ground state.
These configurations are likely to be  very scattered in phase space and so to have zero overlap between themselves. 
 Therefore, in the large $N$ limit  for almost all samples,  $P_{\bf J}(Q)$ has the form
\begin{equation}
P_{\bf J} (Q)= (1-Y_2)\,  \delta(Q) + Y_2 \, \delta(Q-1)
\label{B7}
\end{equation}
where $Y_2$ is the probability of finding at equilibrium two  copies of the system  in the same configuration
\begin{equation}
Y_2=
\frac{\sum_{\bf S} e^{-2 \beta E({\bf S})} }{
\left(\sum_{\bf S} e^{- \beta E({\bf S})} \right)^2 }
\label{B8}
\end{equation}
Similarly
\begin{equation}
Y_k=
\frac{\sum_{\bf S} e^{-k \beta E({\bf S})} }{
\left(\sum_{\bf S} e^{- \beta E({\bf S})} \right)^k }
 \ \ \ \ \  ; 
 \ \ \ \ \   
Y_{k,k'}=
\frac{\sum_{{\bf S}\neq{\bf S}'} 
e^{-k \beta E({\bf S})
-k' \beta E({\bf S}')} }{
\left(\sum_{\bf S} e^{- \beta E({\bf S})} \right)^{k+k'} }
\label{B9}
\end{equation}

 These quantities depend on the sample (i.e. on the realization of the $2^N$ energies $E({\bf S})$).
For the REM, their sample averages  have been calculated in the low temperature \cite{Derrida_1985_Sample}
 (see section 3 for  a simple derivation)
\begin{equation}
 {\langle  Y_k \, Z(\beta)^n    \rangle \over \langle Z(\beta)^n \rangle}  =  {\Gamma(k-\mu)\over \Gamma(1-\mu)}  \ {\Gamma(1-n)  \over \Gamma (k-n) }= \prod_{j=1}^{k-1} {j-\mu \over j-n}
\label{B10}
\end{equation}

\begin{equation}
{\langle Y_{k,k'} \   Z(\beta)^n  \rangle \over \langle Z(\beta)^n \rangle}  =  (\mu-n) {\Gamma(k-\mu)\over \Gamma(1-\mu)}  \ {\Gamma(k'-\mu) \over \Gamma(1-\mu)}   \ {\Gamma(1-n)   \over \Gamma (k+k'-n) }
\label{B11}
\end{equation}
and they agree with the predictions of the replica theory
\cite{Mezard_1984_Nature,Mezard_1984_Replica,Derrida_1997_random,Derrida_2015_Finite}.

Note that setting $n=0$,  the expressions  (\ref{B10}) and (\ref{B11})  reduce to  (\ref{A7}). Also by expanding 
(\ref{B10})  in powers of $n$  one can see that the free-energy and the overlaps are correlated
\begin{equation}
\langle Y_k \log Z(\beta)  \rangle - \langle Y_k \rangle \,  \langle \log Z(\beta) \rangle  = {\Gamma(k-\mu) \over \Gamma(k) \, \Gamma(1 - \mu)} \left( \sum_{q=1}^{k-1} {1 \over q} \right)
\label{B12}
\end{equation}

\subsection{Other distributions of energies}
All the above expressions (\ref{B4},\ref{B6}-\ref{B12})  remain valid  for  more general distributions of energies. For example if the distribution (\ref{B1}) of energies  $E({\bf S})$ is, for large $N$,  of the form
\begin{equation}
P(E) \simeq \sqrt{G''\left({E \over N}\right) \over 2 \pi N} \exp \left[ - N G\left({E \over N}  \right)\right] 
\label{B13}
\end{equation}
where $G$ is a convex function (this would be the case if each energy is the sum of $N$ i.i.d. random variables distributed  according to a continuous distribution), the only change being that the value of $\beta_c$ and the constant $A$ in (\ref{B3}) and (\ref{B5})   become
\begin{equation}
\beta_c=-G'(\epsilon_c)
\ \ \ \ \ ; \ \ \ \ \  A= \sqrt{ G''(\epsilon_c) \over 2 \pi N} \, e^{-N \beta_c \epsilon_c}
\label{B14}
\end{equation}
where $\epsilon_c$ is the minimal solution of $\log(2) - G(\epsilon_c)=0$.

Note, as we will see in section 4, that if the distribution of energies is not of the form (\ref{B13}), in particular when the energies $E({\bf S})$ take only  discrete integer values, the expressions of the overlaps will be quite different.

\subsection{The Poisson process}
Overlaps are non-zero only in the low temperature phase of the REM. This is  why, in this whole paper, we limit our discussion to this low temperature phase where  only the configurations whose energies are close to the  ground state energy matter. The energies of these configurations for a given sample can be  generated by a Poisson process of density
\cite{Derrida_2015_Finite} 
\begin{equation}
\rho(E)= A \,  e^{\beta_c E}
\label{B15}
\end{equation}
 which approximates $2^N P(E)$ in the neighborhood of the ground state energy.  Depending
  on the choice of $P(E)$ in (\ref{B1}) or (\ref{B13}), the amplitude $A$ and $\beta_c$ are given by (\ref{B3}-\ref{B5}) or (\ref{B14}).
In the next section we will see that (\ref{B15})  allows one to recover the predictions (\ref{A7}) of the replica approach.

One can notice that, for the density (\ref{B15}),  the average partition function $\langle Z(\beta)\rangle= \int \rho(E) d E \ \exp[-\beta E] $  is infinite for all values of $\beta$. However, in the low temperature phase $\beta > \beta_c$, (which is the  only temperature range where the overlaps are non-zero), $Z(\beta)$ is finite with probability 1. (In fact it is easy to prove that for all  $C >0$
\begin{equation}
\text{Pro}\Big(Z(\beta)< C\Big) \  >  \
\text{Pro}(Z_1 < C)) \ \text{Pro}(Z_2=0)  \ > \ 
\left(1-{\langle Z_1 \rangle \over C} \right) \ \text{Pro}(Z_2=0) 
\label{B15a}
\end{equation}
where  $Z(\beta) =Z_1+Z_2$ and $Z_1$ represents the contributions to $Z$ of the energies $E > \Lambda$ and $Z_2$  the contributions of the energies  $E<\Lambda$. One has
$$\langle Z_1 \rangle = A\,  {e^{(\beta_c-\beta)\Lambda} \over \beta-\beta_c}      \ \ \ \ \  \text{and}
     \ \ \ \ \  \text{Pro} (Z_2=0)= \exp\left[ -A\,  {e^{\beta_c \Lambda} \over \beta_c} \right] $$
By choosing $\Lambda$ sufficiently negative and $C$ much larger than $\langle Z_1 \rangle$, one can make the r.h.s.  of (\ref{B15a}) as close as needed to 1.) We will see in  section 3 that the Poisson process with the density (\ref{B15}) allows one to recover the known expressions of the negative moments (\ref{B4})  of the REM.
\subsection{Link with sums of random variables with a heavy tail}

In the low temperature phase, the partition function $Z(\beta)$ of the REM can be viewed as a sum of i.i.d. random variables distributed  according to a heavy distribution \cite{Derrida_1997_random}. 
Let $E_1 < E_2 < \dots E_p$ be the $p$ lowest energies of a realization of the Poisson process with density (\ref{B15}). The probability distribution of these energies is
$$\text{Pro}(E_1,\cdots E_p) =A^p \exp\left[\beta_c(E_1+ E_2 + \cdots E_p) - A {e^{\beta_c E_p } \over \beta_c} \right] $$
On the other hand  if one considers a large number $M$ of i.i.d. random positive variables $x_i$ distributed according to a distribution $\rho(x)$ with an heavy tail
$$\rho(x) \sim  B \, x^{-1-\mu} \ \ \ \ \ \text{for large}\ \  x $$   and if one orders these $x_i$'s
the distribution of  $x_1>x_2 \cdots x_p$ is given for large $M$  by
$$P(x_1, \cdots x_p) =M^p \rho(x_1) \rho(x_2) \cdots \rho(x_p) \left[-M\int_{x_p}^\infty  \rho(x) dx \right]
\sim {(M B)^p  \over (x_1 \, x_2 \cdots x_p)^{1+\mu}} \exp\left[-{M B \over \mu \, x_p^{\mu} }\right] $$
We see that the two previous distributions are identical through the change of variables $$
x_i= \left({M B \beta \over A}\right)^{1 \over \mu} e^{-\beta E_i}  \ \ \ \ \ \text{with} \ \ \ \ \ \mu={\beta_c \over \beta}
$$
This shows that the partition function $Z(\beta)= \sum_i e^{-\beta E_i}$ is up to a rescaling a sum of i.i.d. random variables with an heavy tail.
\subsection{The $p$-spin Ising spin glass }
The REM is  the large $p$ limit of the $p$-spin model
\cite{Derrida_1980_Random,Derrida_1981_Random}
  which is a generalisation of the Sherrington-Kirkpatrick model
with an energy given by
\begin{equation}
E_{\bf A} ({\bf S}) = -\sum_{i_1 \le i_2 \cdots i_p} A_{i_1\cdots i_p} \ S_{i_1} \ S_{i_2} \cdots S_{i_p} 
\label{p-spin}
\end{equation}
where the $p$-spin interactions $A_{i_1\cdots i_p}$ are quenched random variables distributed according to

\begin{equation}
\rho(A_{i_1\cdots i_p})= \sqrt{\frac{N^{p-1}}{\pi \,  p!}} \exp\left[ - \frac{(A_{i_1\cdots i_p})^2 \, N^{p-1} }{   p! } \right] 
\label{p-spin1}
\end{equation}
It is easy to see  from (\ref{p-spin},\ref{p-spin1}) that  the energies $E({\bf S})$ are still distributed according to (\ref{B1}) and that their covariances are given   (for large $N$)  by
\begin{equation}
\langle E({\bf S}) \, E({\bf S'}) \rangle  \simeq {N \over 2 } \big[  
q({\bf S},{\bf S'}) \big]^p  
\label{p-spin2}
\end{equation}
Clearly this covariance vanishes in the large $p$ limit when $
|q({\bf S},{\bf S'})| < 1$ and it was argued in 
\cite{Derrida_1980_Random,Derrida_1981_Random}
that in the large $p$ limit, the $p$-spin model should be equivalent to the REM.
Indeed,
using the RSB approach Gross  and M\'{e}zard  \cite{Gross_1984_simplest} were able to calculate  the average free-energy as well as $\langle Y_2 \rangle $ in the large $p$ limit and  their predictions  agree with  (\ref{B3}) and (\ref{A7}) with $\mu$ given by (\ref{B6}). A RSB solution for finite $p$ was then developed by Gardner in 1985 \cite{Gardner_1985_Spin}.

\section{The general framework of a Poisson REM}
In this section we give a few general formulas (\ref{C2}-\ref{C6}) allowing to calculate the moments of the partition and the average overlaps  for a REM whose energies are the points of a Poisson process of an arbitrary density $\rho(E)$.
\subsection{An arbitrary density}
For a Poisson REM of density $\rho(E)$,
 the energies  $E({\bf S})$ are the points of   a Poisson process
of density $\rho(E)$ meaning that in each infinitesimal interval $(E,E+dE)$ with  $dE \ll 1$
there is one configuration 
 with probability $\rho(E) dE$ and with probability $1 - \rho(E) dE$ that there is no configuration at that energy.  
(If the integral of $\rho(E)$ is infinite, then the number of configurations is infinite with probability 1. Then the partition function at some low value of $\beta$  i.e. at high enough temperature could become infinite. However, in all cases we consider here, only the lowest energy levels matter in the low temperature phase implying that the negative moments of the partition function are finite and the overlaps have non trivial values.) 

For each realization of the process, one has a sequence of energies $E_1, E_2, \cdots E_p,  \cdots$ and the partition function is  by definition
\begin{equation}
Z(\beta)= \sum_p e^{-\beta E_p}
\label{AA1}
\end{equation}
Given $\rho(E)$ one can first show that
\begin{equation}
 \langle e^{-t Z(\beta)} \rangle = e^{\phi(t)}
\label{C1}
\end{equation}
where
\begin{equation}
\phi(t)= \int \rho(E) dE \left( \exp\left[-t \, e^{-\beta E}\right]  -1\right)
\label{C2}
\end{equation}
To do so  one simply writes that
\begin{align*}
\langle e^{- t Z(\beta)} \rangle & = 
\prod_E \Big(1- \rho(E) dE  +e^{-t e^{-\beta E}} \, \rho(E) dE \Big) 
\end{align*}
and because $dE \ll 1$ one can exponentiate to obtain (\ref{C1},\ref{C2}).
Knowing $\phi(t)$, one can then write the expressions of integer and non integer moments of the partition function. For example for negative $n$ one has
 
\begin{equation}
\langle Z(\beta)^n \rangle = {1 \over \Gamma(-n)} \int_0^\infty t^{-n-1} dt \, e^{\phi(t)}  \ \ \ \ \  \text{for}  \ \ n<0 
\label{C3}
\end{equation}
In the same manner (see the proof below) one can show  (see (\ref{B9})) that for $n < k$
\begin{equation}
\langle Y_k Z(\beta)^n \rangle = \langle Z(k \beta) \, Z(\beta)^{n-k} \rangle = {1 \over \Gamma(k-n)} \int_0^\infty t^{k-n-1} dt \, \Phi_k(t)\   e^{\phi(t)}  
\label{C4}
\end{equation}

or
\begin{equation}
\langle Y_{k,k'} Z(\beta)^n \rangle = {1 \over \Gamma(k+k'-n)} \int_0^\infty t^{k+k'-n-1} dt \, \Phi_k(t)\ \Phi_{k'}(t)  \    e^{\phi(t)}  
\label{C5}
\end{equation}
where 
\begin{equation}
\Phi_k(t) =  \int \rho(E) dE  \, \exp\left[-k \beta E -t \, e^{-\beta E}\right]  
\label{C6}
\end{equation}
{\it Proof:} For example to obtain (\ref{C4}) one  can use  (see(\ref{B9})) 
the fact that
$$Y_k = {Z(k \beta) \over Z(\beta)^k} $$
and that
$$\langle Z(k\beta) e^{-t Z(\beta)}\rangle = \sum_E  \rho(E) dE e^{-k \beta E-t e^{-\beta E}}  \  
\prod_{E'\ne E} \Big(1- \rho(E') dE'  +e^{-t e^{-\beta E'}} \, \rho(E') dE' \Big)$$ 
Then one can replace the sum over $E$ by an integral and the product as the exponential of an integral (in fact the condition $E\neq E'$  can be forgotten). A similar reasoning leads to (\ref{C5}).

\subsection{The  case of  the exponential density (\ref{B15})}
In this section we show how to recover  (\ref{B4},\ref{B10},\ref{B11})
using (\ref{C1}-\ref{C6}) when the density $\rho(E)$   is exponential.
For the density  given by (\ref{B15}),  the functions $\phi(t)$ and $\Phi_k(t)$   defined in (\ref{C2}) and (\ref{C6}) have explicit exact expressions
in the low temperature phase ($\beta> \beta_c$).
\begin{equation}
\phi(t)={A \over \beta}\  \Gamma\left(-{\beta_c \over \beta} \right) \, t^{\beta_c \over \beta}
\ \ \ \ \ ; \ \ \ \ \
\Phi_k(t)={A \over \beta}\  \Gamma\left(k-{\beta_c \over \beta} \right) \, t^{{\beta_c \over \beta}-k}
\label{C7}
\end{equation}
(In the high temperature phase, i.e. for $\beta< \beta_c$, it turns out that $\phi(t)$ is infinite for the density (\ref{B15}). There it is no longer appropriate to replace the original REM by a Poisson process with the exponential density (\ref{B15}).) Using (\ref{C7}) the integrals in (\ref{C3}-\ref{C5}) can be computed exactly. One finds

\begin{equation}
  \left\langle Z(\beta)^n \right\rangle =\left(- {A \over \beta}\,  \Gamma\left(-{\beta_c \over \beta} \right)\right)^{n {\beta \over \beta_c}} \ {\Gamma(1- n {\beta \over \beta_c}) \over \Gamma(1-n)}
\label{C8}
\end{equation}
which is identical to (\ref{B4}) with the definition (\ref{B6}) of $\mu$. One also recovers that way   from (\ref{C4},\ref{C5}) the expressions (\ref{B10},\ref{B11}) of the overlaps.

\section{The double exponential and its consequences} 
One simple case for which the expressions (\ref{B10}) and (\ref{B11}) of the overlaps are no longer valid is when the density $\rho(E)$ is a sum of two exponentials. Here we show that one needs to let fluctuate the parameter $\mu$ in these expressions.
The generalization to  the sum of an arbitrary number of exponentials will be  straightforward. This will allow us to calculate   to recover, in a much easier way,  earlier results on finite size corrections or on the effect of  discrete energies

\subsection{The double exponential case}
In this case, the density $\rho(E)$ is 
\begin{equation}
\rho(E) = A_1 \, e^{\beta_1 \, E}
+  A_2 \, e^{\beta_2 \, E}
\label{D1}
\end{equation}
Then (\ref{C2}) and (\ref{C6}) become
\begin{equation}
\phi(t)  =  B_1 \, t^{\mu_1} \, \Gamma(-\mu_1)  +  B_2 \, t^{\mu_2} \, \Gamma(-\mu_2)   
\label{D2}
\end{equation}
\begin{equation}
\Phi_k(t)  =  B_1 \, t^{\mu_1-k} \, \Gamma(k-\mu_1)  +  B_2 \, t^{\mu_2-k} \, \Gamma(k-\mu_2)  
\label{D3}
\end{equation}
with
\begin{equation}
B_i= {A_i \over \beta} \ \ \ \ \ \text{and}  \ \ \ \ \  \mu_i={\beta_i \over \beta} 
\label{D4}
\end{equation}

The relation  (\ref{C3}) (after an integration by parts)  can be written as
\begin{equation}
 \langle Z(\beta)^n \rangle = {-1 \over \Gamma(1-n)} \, \int_0^\infty t^{-n} dt  \, \phi'(t) \, e^{\phi(t)}
\label{D5}
\end{equation}
Then by 
replacing $\phi'(t)$  using (\ref{D2}) and using again (\ref{C3}), one gets for $n <0$
$$
\langle Z(\beta)^n \rangle =
  B_1 \,  {\Gamma(1-\mu_1)\, \Gamma(\mu_1-n)
 \over \Gamma(1-n)}
 \, \langle Z(\beta)^{n-\mu_1}\rangle 
+   B_2 \,  {\Gamma(1-\mu_2)\, \Gamma(\mu_2-n)
 \over \Gamma(1-n)}
\, \langle Z(\beta)^{n-\mu_2} \rangle
$$
which can be written as
\begin{equation}
\langle Z(\beta)^n \rangle = \sum_{i=1,2}
  B_i \  {\Gamma(1-\mu_i)\, \Gamma(\mu_i-n)
 \over \Gamma(1-n)}
 \, \langle Z(\beta)^{n-\mu_i}\rangle 
\label{D6}
\end{equation}

For the overlaps, starting from (\ref{C4}) with $\Phi_k(t)$ given by (\ref{D3})  one can  repeat the same procedure to get
\begin{equation}
\langle Y_k\ Z(\beta)^n \rangle =
   \int_0^\infty t^{k-n-1} dt  
  \left[\sum_{i=1,2} B_i \ {\Gamma(k-\mu_i)  \over \Gamma(k-n)}
\, t^{\mu_i-k} 
 \,  e^{\phi(t)} \right]    
\label{D6a}
\end{equation}
which gives 
\begin{equation}
\langle Y_k\ Z(\beta)^n \rangle = \sum_{i=1,2} 
  B_i \, { \Gamma(k-\mu_i) 
\ \Gamma(\mu_i-n)
\over \Gamma(k-n)}
\, \langle Z(\beta)^{n-\mu_i}\rangle  
\label{D7}
\end{equation}
Using  (\ref{D6}) 
 one can rewrite 
 (\ref{D7}) as
\begin{equation}
{\langle Y_k\ Z(\beta)^n \rangle \over \langle  Z(\beta)^n \rangle} =
 \sum_{i=1,2} 
{
\Gamma(k-\mu_i) \over 
 \Gamma(1-\mu_i)}  \ 
{\Gamma(1-n)  
\over 
 \Gamma(k-n)
}
\ W_i
\label{D8}
\end{equation}
where  the weights $W_i$ are given by
\begin{equation}
W_i =
  B_i \  {\Gamma(1-\mu_i)\, \Gamma(\mu_i-n)
 \over \Gamma(1-n)}
 \, {\langle Z(\beta)^{n-\mu_i}\rangle 
 \over \langle Z(\beta)^n \rangle }
\label{D9}
\end{equation}
We see that  the average overlaps in (\ref{D8}) have the same expression as in   (\ref{B10}) except that now,  for the double exponential (\ref{D1}), the parameter $\mu$ fluctuates between  two values $\mu_1$ or $\mu_2$.

The averages of other overlap functions can be  derived  from (\ref{C5}) in the same way: for example using the fact that 
 by iterating (\ref{D6}) one has
\begin{equation}
\langle Z(\beta)^n \rangle =
\sum_{i=1,2} \sum_{j=1,2}
  B_i \, B_j \, { \Gamma(1-\mu_i) \Gamma(1-\mu_j) \, \Gamma(\mu_i+ \mu_j-n) \over (\mu_i-n) \,  \Gamma(1-n)} \langle Z(\beta)^{n-\mu_i-\mu_j}\rangle 
\label{D10}
\end{equation}
and one gets, using (\ref{D3}) twice in (\ref{C5})
\begin{equation}
{ \langle Y_{k,k'} Z(\beta)^n \rangle \over \langle Z(\beta)^n \rangle}  =
\sum_{i=1,2}  
\sum_{j=1,2} (\mu_i-n) 
{ \Gamma(k-\mu_i)   \over \Gamma(1-\mu_i) } 
{ \Gamma(k'-\mu_j)   \over \Gamma(1-\mu_j) } 
{\Gamma(1-n) \over \Gamma(k+k'-n)} 
 W_{i,j}
\label{D11}
\end{equation}
with the weights $W_{i,j}$ given by
\begin{equation}
 W_{i,j}=
  B_i \, B_j \, { \Gamma(1-\mu_i) \Gamma(1-\mu_j) \, \Gamma(\mu_i+ \mu_j-n) \over (\mu_i-n) \,  \Gamma(1-n)}
{ \langle Z(\beta)^{n-\mu_i-\mu_j}\rangle  \over \langle Z(\beta)^n \rangle} 
\label{D12}
\end{equation}
Clearly (\ref{D11}) is a generalization of (\ref{B11})   where the $\mu$'s fluctuate. We see in (\ref{D12}) that the $\mu_i$'s are correlated (as $W_{i,j} \ne W_i \, W_j$).
Note that although the $W_{i,j}$  is not symmetric under  the exchange $\mu_i \leftrightarrow \mu_j$,  the symmetry is restored in the sum (\ref{D11}).
 \\ \ \\
{\it Remark:} One possible realization of the  double exponential density (\ref{D1}) would be to consider a REM with  a total of $\alpha_1^N + \alpha_2^N $ configurations,  the first $\alpha_1^N$ configurations having energies distributed according to  $P_1(E) = \exp[-E^2 /(N a_1)] / \sqrt{\pi a_1}$ and the last $\alpha_2^N$ according to $P_2(E) = \exp[-E^2 /(N a_2)] / \sqrt{\pi a_2}$.  When the parameters $\alpha_1,\alpha_2,a_1,a_2$ satisfy the  following condition
$$ \sqrt{a_1 \log \alpha_1}=  \sqrt{a_2 \log \alpha_2} \equiv -\epsilon$$ the  density 
$\rho(E)$ near the ground state energy becomes
$$\rho(E) =
{1 \over \sqrt{\pi N a_1} } \exp\left[ 2 \sqrt{\log \alpha_1 \over a_1}( E- N \epsilon) \right]
+ {1 \over \sqrt{\pi N a_2} } \exp\left[ 2 \sqrt{\log \alpha_2 \over a_2}( E- N \epsilon) \right]
$$
which is indeed of the form (\ref{D1}).
For more general spin glass models, one can imagine that the double exponential density (\ref{D1})  would also be relevant  near a first order  transition between two low temperature phases  that exhibit  a one step RSB.
\subsection{The replica approach in the case of the double exponential}
\begin{figure}[h]
 \centerline{\includegraphics[width=6.cm]{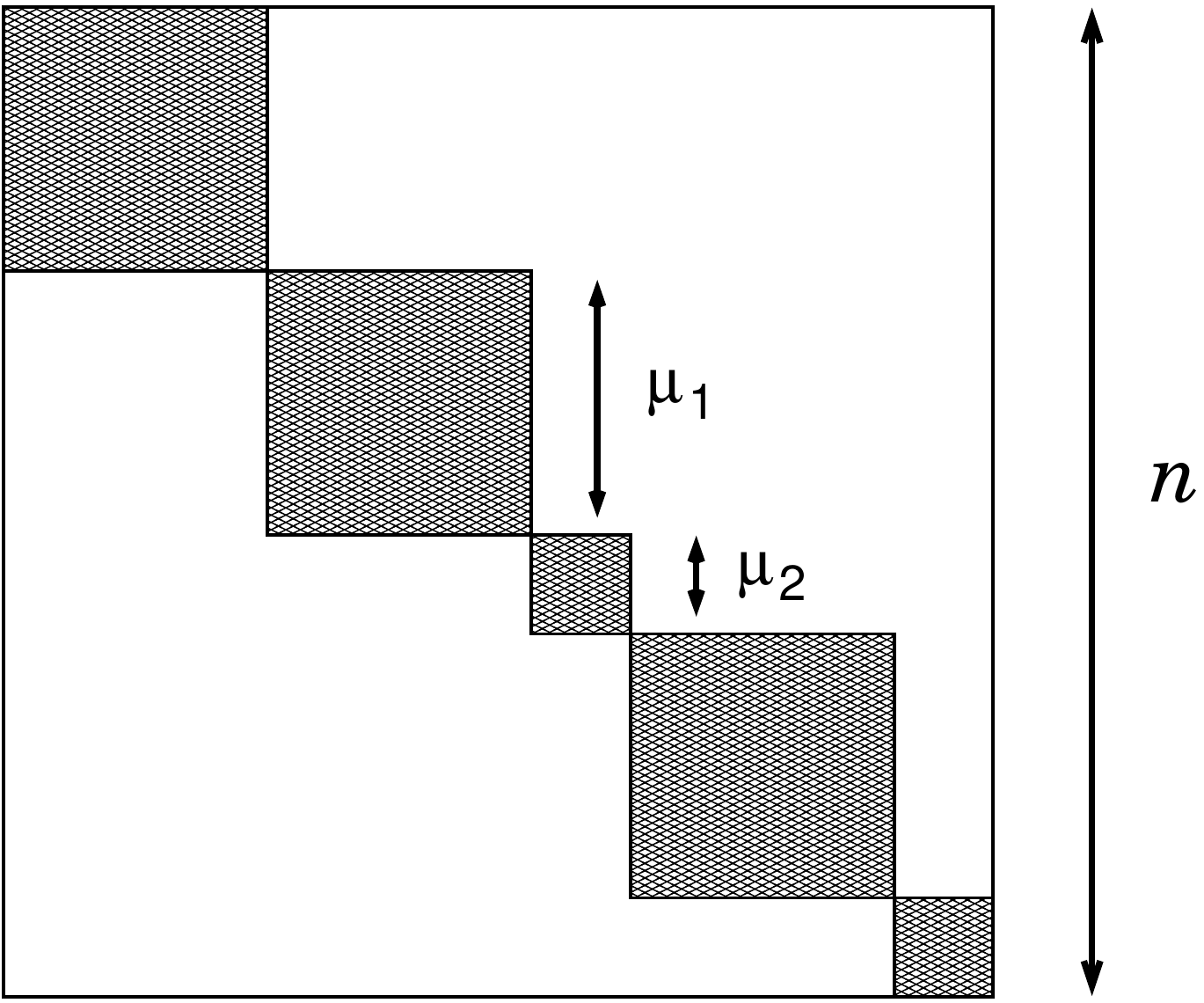}}
 \caption{In the case of the double exponential, the overlap matrix of Figure \ref{peter-fig1} shoud be replaced by a matrix whose block sizes fluctuate. }
\label{peter-fig2}
\end{figure}

In the appendix A, we recall how to obtain the expressions (\ref{A7},\ref{B10},\ref{B11}) using the Parisi matrix shown in Figure  \ref{peter-fig1}. 
We are now going to see that in order to  recover the above expressions (\ref{D8},\ref{D11}) one needs to consider  matrices of the form shown in Figure \ref{peter-fig2}
and to average over matrices of this shape by letting the number $n_i$ of blocks of size $\mu_i$ to fluctuate.
Let us imagine that a  matrix of overlaps is a $n \times n$ matrix, as in Figure \ref{peter-fig2}, where the diagonal blocks have  varying sizes $\nu_\alpha$ so that
\begin{equation}
n= \sum_\alpha   \nu_\alpha
\label{D13}
\end{equation}
The probability that $k$  different replicas chosen among $n$ replicas belong to the same block is
\begin{equation}
Y_k= \sum_{\alpha}  {\nu_\alpha ! \over (\nu_\alpha-k) ! } \, {(n-k)! \over n!} 
\label{D14}
\end{equation}
These expressions depend on the number $n$ of replicas and on the sizes $\nu_\alpha$ of the blocks.
Now if we allow  $n$ and the $\nu_\alpha$'s to take non-integer values, as in the original RSB scheme \cite{Parisi_1979_Infinite} this can be written as
\begin{equation}
Y_k= \sum_{\alpha} 
 {\Gamma(k-\nu_\alpha)  \over \Gamma(-\nu_\alpha) } \, {\Gamma(-n) \over \Gamma(k-n)} 
\label{D15}
\end{equation}
Similarly
\begin{equation}
Y_{k,k'}= \sum_{\alpha} \sum_{\alpha' \neq \alpha} 
 {\Gamma(k-\nu_\alpha)  \over \Gamma(-\nu_\alpha)  } \,
 {\Gamma(k'-\nu_{\alpha'})  \over \Gamma(-\nu_{\alpha'})  } \,
 {\Gamma(-n) \over \Gamma(k+k'-n)} 
\label{D16}
\end{equation}

Let us imagine that the $\nu_\alpha$ can take possible values $\mu_i$ (in the case of two exponentials there are two possible values $\mu_1$ and $\mu_2$ defined in (\ref{D4})) and let $n_i$ be the number of $\nu_\alpha$'s taking the value $\mu_i$.  Then (\ref{D13}) becomes
\begin{equation}
\sum_i n_i \, \mu_i=n 
\label{D16a}
\end{equation}
and going from one matrix to the other the values of the $n_i$ fluctuate (while the $\mu_i$'s don't) keeping this sum constant.

Averaging over the $n_i$'s one gets
$${\langle Y_k \, 
Z(\beta)^n \rangle  \over \langle Z(\beta)^n \rangle  }
 = \sum_i \langle n_i \rangle \, {\Gamma(k-\mu_i)  \over \Gamma(-\mu_i)  } \, 
{\Gamma(-n) \over \Gamma(k-n)} 
 = \sum_i {\langle n_i \rangle \,  \mu_i \over n}  \ {\Gamma(k-\mu_i)  \over \Gamma(1-\mu_i)  } \, {\Gamma(1-n) \over \Gamma(k-n)} 
$$
which has the form (\ref{D8})  if one identifies 
\begin{equation}
W_i ={\langle n_i \,\rangle \mu_i \over n} 
\label{D17}
\end{equation}
One can   interpret   the weight  $W_i$ as  the probability that one among $n$ replicas belongs to a block of size $\mu_i$.

The matrix in Figure \ref{peter-fig2} is characterized by the numbers $n_1$ and $n_2$ of blocks of size $\mu_1$ and $\mu_2$. The sizes $\mu_1$ and $\mu_2$ of the blocks are fixed.   Only $n_1$ and $n_2$ fluctuate with the constraint (\ref{D16a}) and to recover (\ref{D8},\ref{D11}) one needs to average over $n_1$ and $n_2$. Note that in this picture, although $\mu_1$ and $\mu_2$ are simply given by (\ref{D4}),  the expressions of the weights $W_i$ and $W_{i,j}$ in (\ref{D9},\ref{D12}) are less explicit as they require the knowledge of negative moments of the partition function.

Similarly 
\begin{multline}
{\langle Y_{k,k'} \,      
Z(\beta)^n \rangle  \over \langle Z(\beta)^n \rangle  }
 =\left[ \sum_i \langle n_i^2- n_i\rangle \ 
{\Gamma(k-\mu_i)  \over \Gamma(-\mu_i)  } 
{\Gamma(k'-\mu_i)  \over \Gamma(-\mu_i)  } \right. \\
 \left. + \sum_i  \sum_{j \neq i} \langle n_i n_j\rangle \ 
{\Gamma(k-\mu_i)  \over \Gamma(-\mu_i)  } 
{\Gamma(k'-\mu_j)  \over \Gamma(-\mu_j)  } \,
\right] 
 {\Gamma(-n) \over \Gamma(k+k'-n)} 
\label{D17a}
\end{multline}
which has the form (\ref{D11})  if one identifies  
\begin{equation}
W_{i,i} ={\langle n_i (n_i-1)\rangle\,  \mu_i^2\over n(n-\mu_i)}  \ \ \ \ \ \text{ and for $i\neq j$} \ \ \ \ \
W_{i,j} ={\langle n_i n_j\rangle \,  \mu_i \mu_j\over n(n-\mu_i)} 
\label{D18}
\end{equation}

In all cases i.e. whether $i=j$ or $i\neq j$, one can  interpret  $W_{i,j}$ as the probability that choosing 2 replicas belonging to different blocks, the first one is in a block of size $\mu_i$ and the  second one in a block of size $\mu_j$.

\subsection{The finite size corrections to the REM}
It is straightforward to generalize the above results for the double exponential (\ref{D1}) to the sum of an arbitrary  number of exponentials
\begin{equation}
\rho(E) =\sum_j A_j \ e^{\beta_j E}
\label{D19}
\end{equation}
The density of energies (\ref{B1}) of the REM near the  ground state energy  can be written as  

\begin{equation}
{2^N \over \sqrt{\pi N}} \exp \left[-{E^2 \over N} \right] =  \int {dy \over   \pi  \, N }
e^{- N y^2} \exp\left[  (\beta_c + 2 i y )(E+ N \sqrt{\log 2} )\right]  \label{D20}
\end{equation}
which is of the form   (\ref{D19}) if one allows the $\beta_j$'s  to take  small imaginary values $\beta_c + 2 iy $. Therefore as in the double exponential case, this implies that  block sizes $\mu_j $ (see (\ref{D4}))  can become complex.

Because the possible values of the block   sizes $\mu_j= {\beta_c + 2 i y \over \beta}$  are now complex (see (\ref{D4})),   the weights $W_i$  and the numbers $n_i$ may  be complex too. Using the fact that $y$ is distributed as in (\ref{D20}), one gets that for large $N$
$$\langle \mu_j^2 \rangle - \langle \mu_j \rangle^2 \sim  -{2 \over N \beta^2} $$
recovering the prediction in equation (55) of \cite{Derrida_2015_Finite} that the block size have a negative variance.
(Note that the variance is negative simply because the box sizes $\mu_i$ are complex).

\subsection{The case of discrete energies}
The case where the energies take only discrete values can also be treated the same way, for example if the energies $E({\bf S})$ were sums of $N$ random  numbers taking  only integer values \cite{Derrida_2022_discrete,Ogure_2005_exact,Bouchaud_2003_Energy} In that case the density  $\rho(E)$ near the ground state energy could be written as
\begin{equation}
\rho(E) =A \ e^{\beta_c (E } \sum_{n=-\infty}^\infty \delta(E-n)= \sum_{p=-\infty}^\infty  A \  e^{(\beta_c + 2 i \pi  p)E} 
\label{D21}
\end{equation}

In this case too, the density $\rho(E)$ is a particular case of (\ref{D19}). Therefore  one can use the above formulas allowing the  $\beta_j$ to take complex values $\beta_c + 2 i \pi p$. The block sizes $\mu_i$'s then  take  complex values. 
Therefore according to (\ref{D9},\ref{D17}) the replica interpretation is  the "probabilities"  or rather the $W_p$ (which do not need here to be  positive or even real) that a replica belongs to a block of size $\mu_p={\beta_c +2i \pi p \over \beta}$ are given by
\begin{equation}
W_p =
 {A\over \mu_p} \  {\Gamma(1-\mu_p)\, \Gamma(\mu_p-n)
 \over \Gamma(1-n)}
 \, {\langle Z(\beta)^{n-\mu_p}\rangle 
 \over \langle Z(\beta)^n \rangle }
\ \ \ \ \ \ \text{with}  \ \ \ \ \mu_p={\beta_c +2i \pi p \over \beta}
\label{D22}
\end{equation}

\begin{figure}[h]
 \centerline{
\includegraphics[width=10.5cm]{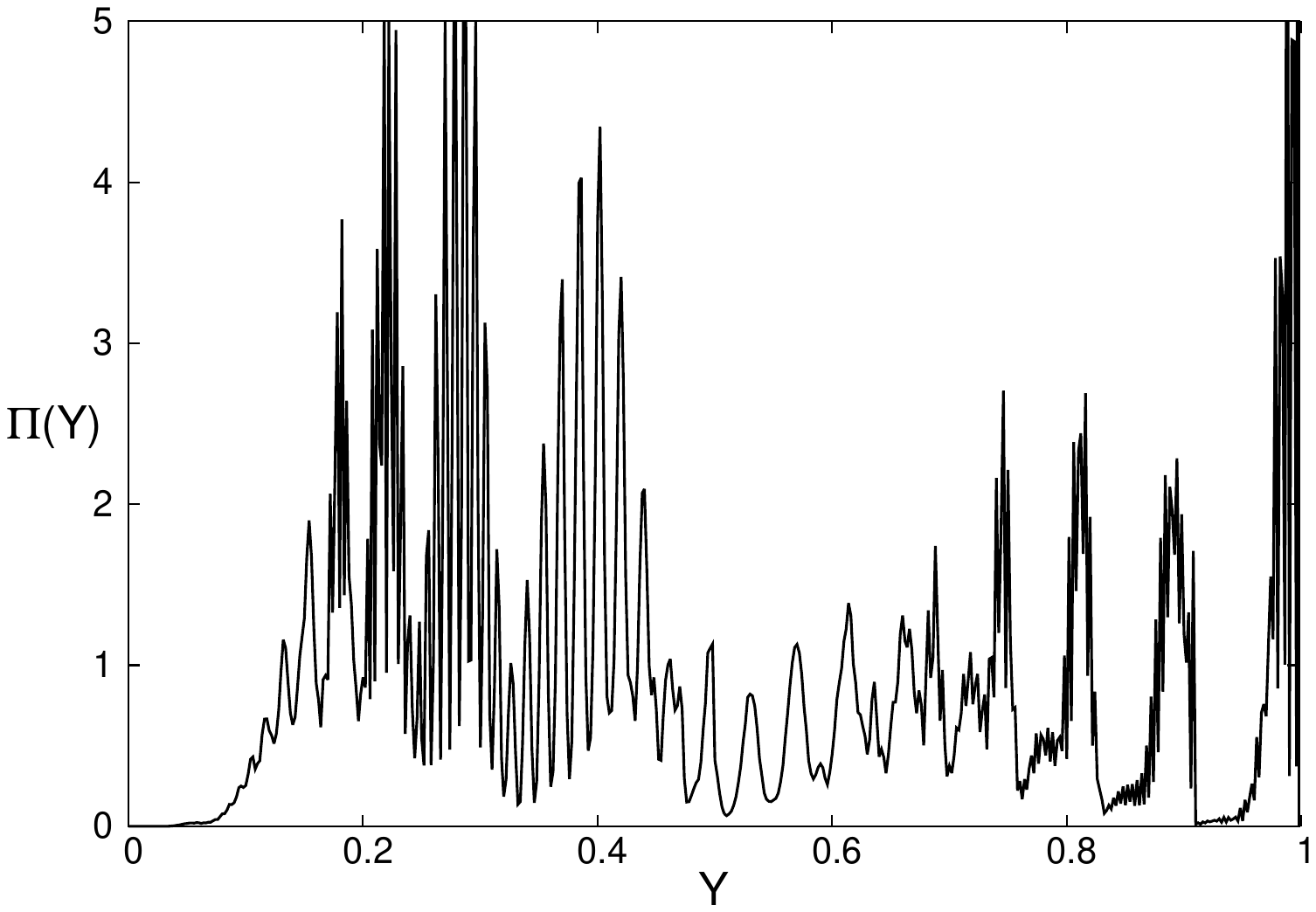}}
 \caption{The distribution $\Pi_\mu(Y)$ when the energies take only integer values. Here like in Figure \ref{figure1},  $\beta=3 \beta_c$. }
 \label{peter-fig4}
 \end{figure}

One  way of illustrating that the statistics  of the overlaps are strongly  modified is to look at the  distribution $\Pi_\mu(Y)$ drawn  in Figure \ref{peter-fig4} and to compare it to the shape of Figure \ref{figure1} for continuous energies.

It is easy to see that changing  the amplitude $A$ in (\ref{D21}) to $A e^{\beta_c}$ has the effect of shifting all the energies by $1$ implying that $\Pi_\mu$ and the overlaps remain unchanged. Other changes of $A$ (for example $A \to A e^{\beta_c x }$  when $x$  is not an integer change the shape of the distribution $\Pi_\mu (Y)$ and the average overlaps $Y_k$ with a periodic dependence on the parameter $x$ (as observed in \cite{Derrida_2022_discrete}).

\section{The overlaps of the REM at two temperatures}

Considering the overlaps between configurations at different temperatures in the same energy landscape is a way to study temperature chaos \cite{Bray_1987_Chaotic}  (for a review see \cite{Rizzo_2009_Chaos}). It is well known that the REM as well as other models, such as directed polymers in a random medium (in their tree version), do not exhibit temperature chaos \cite{Sales_2001_Rejuvenation,Krzakala_2002_Chaotic,Kurkova_2003_Temperature,Pain_2021_Two}. Still it is  interesting to understand how the replica approach works to predict these overlaps. A detailed analysis of this question is given in \cite{Derrida_2021_One}. In this section we provide an alternative analysis extending the replica-free calculation of the previous section to the two temperature case and obtaining an expression for the overlaps that gives an explicit formula in terms of the probabilities of fluctuating block sizes. As in the case of finite size corrections or of discrete energies (see section 4)  we will see that, in the two temperature case, the block sizes take complex values.

Let us consider the overlaps between  $k$ configurations at inverse temperature $ \beta$ and $k'$  
configurations at inverse temperature $\beta'$.   For a given sample, the probability $U_{k,k'}$ that all these $k+k'$ configurations occupy the same energy level is
\begin{equation}
U_{k,k'}= {\sum_{\bf S} e^{-(k\beta +k'\beta') E({\bf S})} \over Z(\beta)^k \ Z(\beta')^{k'}} 
={Z(k\beta +k'\beta') \over Z(\beta)^k \ Z(\beta')^{k'}} 
\label{F0}
\end{equation}
and our goal is to determine the following sample averages 
 $\langle Z(\beta)^n Z(\beta')^{n'} \rangle$
and 
 $\langle U_{k,k'}  Z(\beta)^n Z(\beta')^{n'}\rangle $ in the low temperature phase.
Although there is a perfect symmetry  under the exchange 
$(\beta,k,n) \leftrightarrow 
(\beta',k',n')$, this symmetry is not apparent in some expressions below, but in fact it is indeed respected.

For negative $n$ and $n'$, one can repeat what we did in (\ref{C1}-\ref{C3}) and write
\begin{equation}
\langle Z(\beta)^n \, Z(\beta')^{n'} \rangle = {1 \over \Gamma(-n) \, \Gamma(-n')} \int_0^\infty t^{-n-1} dt \int_0^\infty u^{-n'-1} du \ e^{\varphi(t,u)}
\label{F1}
\end{equation}
where 
\begin{equation}
\varphi(t,u) = \int \rho(E) dE \left(\exp\left[ -t e^{-\beta E} - u e^{-\beta' E} \right] -1 \right)
\label{F2}
\end{equation}
One can rewrite (\ref{F1})  as 
\begin{equation}
\langle Z(\beta)^n \, Z(\beta')^{n'} \rangle = {-1 \over \Gamma(1-n) \, \Gamma(-n')} \int_0^\infty t^{-n} dt \int_0^\infty u^{-n'-1} du \ e^{\varphi(t,u)} \ {d \varphi(t,u) \over dt} 
\label{F3}
\end{equation}
We are now going to use  the following  complex integral representation (called the Cahen-Mellin representation, see for example \cite{Paris_2001_Asymptotics}) of the exponential
\begin{equation}
e^{-t} = \int_{-\infty}^\infty {dy \over 2 \pi} \  \Gamma(\nu+i y) \ t^{-\nu-i y} \ 
\label{F4}
\end{equation}
where $\nu$ is real and  $ \nu > 0$.  (In (\ref{F4}) the integral does not depend on $\nu$ as long as $\nu >0$ simply because the integral path can be deformed without crossing any singulatity.)

For the exponential density (\ref{B15}) and transforming $\exp\left[ - t e^{-\beta E} \right]$ according to (\ref{F4})   one gets, (note that in the low temperature phase $\beta>\beta_c$ and $\beta'>\beta_c$)
$${d \varphi(t,u) \over dt} 
=  -A \int  dE e^{(\beta_c  -\beta)E} 
 \exp\left[ -u e^{-\beta' E} \right]   \int_{-\infty}^\infty {dy \over 2 \pi} \  \Gamma(\nu+i y) \ t^{-\nu-i y}
 e^{( \nu + i y)\beta E}
$$
which, after integrating over $E$, becomes
\begin{equation}
{d \varphi(t,u) \over dt} = -
{A \over \beta'}
 \int_{-\infty}^\infty {dy \over 2 \pi} \  \Gamma(\nu+i y)  \ 
\Gamma\left( {\beta(1-\nu-i y) -\beta_c \over \beta'}\right)  
  t^{-\nu-i y}  \ 
u^{  \beta_c -\beta (1-\nu -  i y) \over \beta'}     
\label{F5}
\end{equation}
Then one gets from (\ref{F3}) using (\ref{F1})
\begin{multline*}
\langle Z(\beta)^n \, Z(\beta')^{n'} \rangle = 
{A \over \beta'}
 \int_{-\infty}^\infty {dy \over 2 \pi} \  {\Gamma(\nu+i y)  \ 
\Gamma\left( {\beta(1-\nu-i y) -\beta_c \over \beta'}\right)  
 \over \Gamma(1-n) \, \Gamma(-n')} 
 \Gamma( 1-n-\nu - i y) \\
 \times
 \Gamma\left( {-n'+{\beta_c -\beta(1-\nu-i y\over \beta'}}\right)
\left\langle
Z(\beta)^{n-1+\nu+i y}
 Z(\beta')^{n'-{\beta_c\over \beta'} +{\beta(1-\nu-iy)\over \beta'}}
\right\rangle 
\end{multline*}
in other words
\begin{multline}
\langle Z(\beta)^n \, Z(\beta')^{n'} \rangle = 
{A \over \beta'}
 \int_{-\infty}^\infty {dy \over 2 \pi} \  {\Gamma(1-\mu_y )  \ 
\Gamma\left( -\mu'_y\right)  
 \over \Gamma(1-n) \, \Gamma(-n')} 
 \Gamma( \mu_y-n)
 \Gamma\left( \mu'_y-n'\right)
\left\langle
Z(\beta)^{n-\mu_y}
 Z(\beta')^{n'-\mu'_y}
\right\rangle 
\label{F6}
\end{multline}
where 
\begin{equation}
\mu_y= 1-\nu - i y 
\ \ \ \ \  \text{and} \ \ \ \ \ 
\mu'_y = {\beta_c- \beta \mu_y \over \beta'}
\label{F7}
\end{equation}
(We have checked (\ref{F6}) against the explicit expression in \cite{Derrida_2021_One} and we have checked that when $\beta=\beta'$ we recover (\ref{C8}).) Comparing with (\ref{D6}) we see that here  the $n$ replicas  at inverse temperature $\beta$ and the  are grouped in blocks of sizes $\mu_y$ and the  $n'$ at inverse temperature $\beta'$ in blocks of sizes $\mu'_y$. Each block of size $\mu_y$ is associated to a block of size $\mu'_y$ which satisfy
\begin{equation}
\beta \mu_y + \beta' \mu'_y= \beta_c
\label{F8}
\end{equation}
So the  $\mu_y$ and $\mu'_y$  fluctuate but the relation (\ref{F8}) remains fixed. The imaginary part of the block sizes fluctuates between $-i \,\infty$ and $+i \,\infty$, but the real part can be arbitrarily fixed within the constraints
\begin{equation}
  n<\Re(\mu_y)<1 \; , \; n'<\Re(\mu'_y)<0
\label{F9}
\end{equation}
which is a result of the requirement for $\nu>0$ in (\ref{F4}) and that the gamma functions in (\ref{F6}) must have positive arguments. Combining (\ref{F8}) and (\ref{F9}) we obtain
\begin{equation}
    \frac{\beta_c}{\beta} < \Re(\mu_y) < \min \left(1, \frac{\beta_c-\beta' n'}{\beta}\right).
\label{F10}
\end{equation}
(Remember that $n' <0$).

We now turn to the two temperature overlaps defined in (\ref{F0}). With negative $n,n'$ and positive integer $k,k'$ we can use the same approach as (\ref{C4}) to obtain
\begin{equation}
\langle U_{k,k'} \, Z(\beta)^n \, Z(\beta')^{n'} \rangle =
   {1 \over \Gamma(k-n) \, \Gamma(k'-n')} \int_0^\infty t^{k-n-1} dt \int_0^\infty u^{k'-n'-1} du \, \Phi_{k,k'}(t,u) \ e^{\varphi(t,u)}
\label{F11}
\end{equation}
where
\begin{equation}
  \Phi_{k,k'}(t,u) = \int \rho(E) dE  \exp\left[-\beta k E -\beta' k' E -t e^{-\beta E} - u e^{-\beta' E} \right] .
\label{F12}
\end{equation}
Using the Cahen-Mellin representation (\ref{F4}) we can write this as a contour integral and for exponential disorder (\ref{B15}) this simplifies to 
\begin{equation}
\Phi_{k,k'}(t,u) =
  {A \over \beta'}
   \int_{-\infty}^\infty {dy \over 2 \pi} \  \Gamma(\nu+i y)  \ 
  \Gamma\left(k'+ {\beta(k-\nu-i y) -\beta_c \over \beta'}\right)  
    t^{-\nu-i y}  \ 
  u^{ -k'- {\beta(k-\nu-i y) -\beta_c \over \beta'}}     .
\label{F13}
\end{equation}
Substituting in (\ref{F11}) and using (\ref{F1}) gives
\begin{multline}
  \langle U_{k,k'} \, Z(\beta)^n \, Z(\beta')^{n'} \rangle =
     {A \over \beta'}
      \int_{-\infty}^\infty {dy \over 2 \pi} \  
      \frac{\Gamma(k-\mu_y) \ \Gamma\left(k'-\mu'_y\right)}{\Gamma(k-n) \ \Gamma(k'-n')}  \\
  \times    \Gamma(\mu_y-n) \ \Gamma\left(\mu'_y -n'\right)
      \left\langle Z(\beta)^{n-\mu_y} Z(\beta')^{n'-\mu'_y} \right\rangle 
\label{F15}
\end{multline}
where $\mu_y$ and $\mu'_y$ are defined in (\ref{F7}) and must satisfy (\ref{F8}) and (\ref{F10}).

We can express the overlaps in terms of a normalised weight function $W(y)$ similar in concept to formula (\ref{D11})
\begin{equation}
\frac{\langle U_{k,k'} \, Z(\beta)^n \, Z(\beta')^{n'} \rangle}{\langle Z(\beta)^n \, Z(\beta')^{n'} \rangle}=
\int_{-\infty}^\infty dy \  
\frac{\Gamma(k-\mu_y) \ \Gamma\left(k'-\mu'_y\right)}{\Gamma(1-\mu_y) \ \Gamma\left(-\mu'_y\right)} 
\frac{\Gamma(1-n) \ \Gamma(-n')}{\Gamma(k-n) \ \Gamma(k'-n')} \, W(y)
\label{F17}
\end{equation}
with
\begin{equation}
  W(y)= \frac{A}{2 \pi \,\beta'}{\Gamma(1-\mu_y )  \ 
  \Gamma\left( -\mu'_y\right)  
   \over \Gamma(1-n) \, \Gamma(-n')} 
   \Gamma( \mu_y-n)
   \Gamma\left( \mu'_y-n'\right) 
  \frac{\left\langle  Z(\beta)^{n-\mu_y}  Z(\beta')^{n'-\mu'_y}   \right\rangle}{ \langle Z(\beta)^n \, Z(\beta')^{n'} \rangle}.
\label{F18} 
\end{equation}

The weights $W(y)$ can be thought of as the probability that, for a given block, the block size is $\mu_y$ at inverse temperature $\beta$ and $\mu'_y$ at inverse temperature $\beta'$ subject to the condition (\ref{F8}). However, they are complex and there is some freedom, due the choice of $\nu$ in (\ref{F4}), in the choice of where the contour crosses the real axis. Despite this lack of uniqueness we expect that the different choices of weights will lead to the same moments for $\mu_y$ and $\mu'_y$. 

The above expressions (\ref{F6}-\ref{F10}) are clearly non-symmetric under the exchange $(n,\beta,\mu_y) \leftrightarrow (n',\beta',\mu_y')$ . We believe that very much like for the choice of the real part of $\mu_y$, there are many different choices of the distribution of the $\mu_y$ and of the $\mu_y'$ which lead to exactly the same predictions for the moments $\langle Z^n(\beta) Z^{n'}(\beta') \rangle $ or of the average overlaps even though the expressions look different. 

\begin{figure}[h]
  \centerline{\includegraphics[width=9cm]{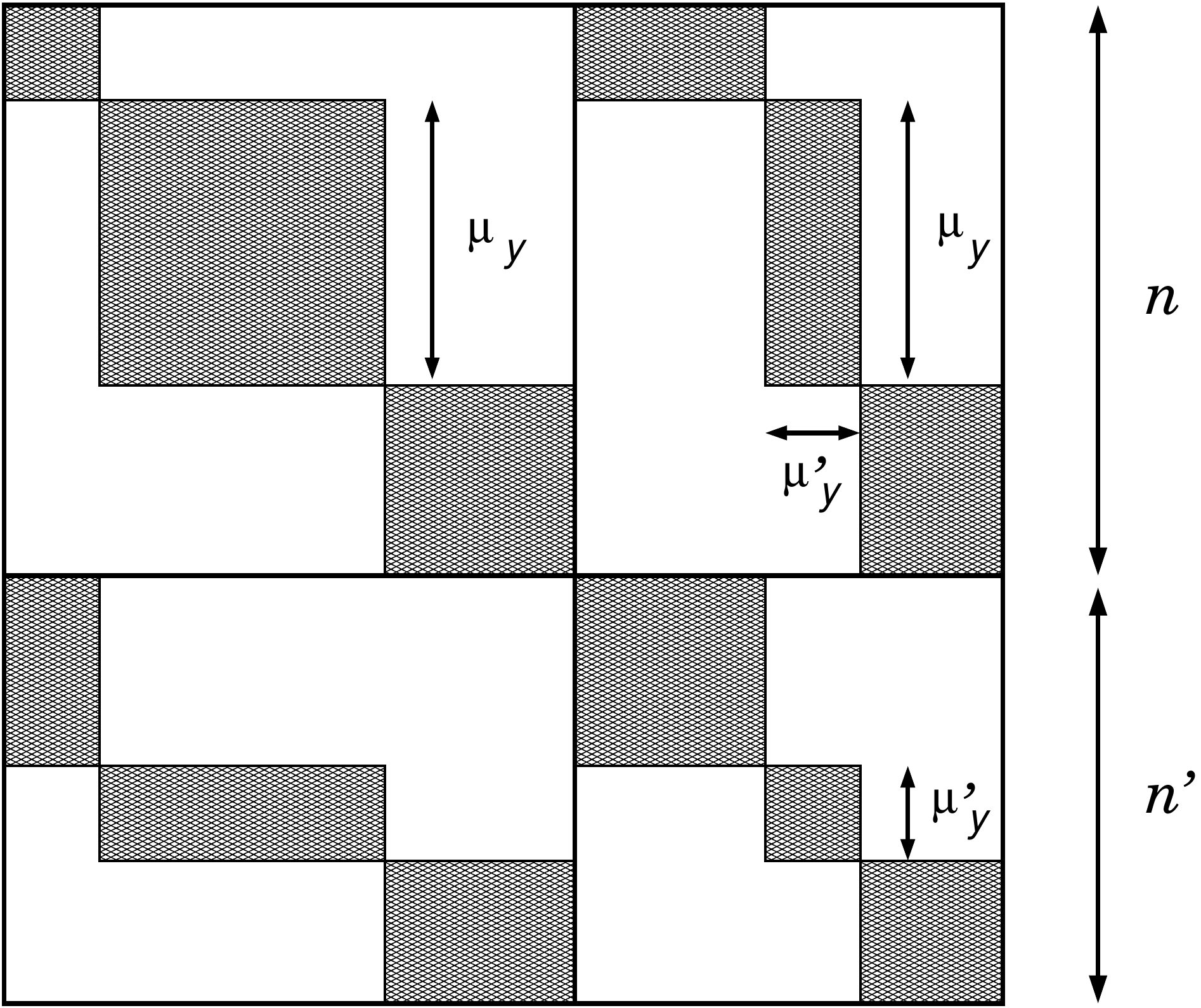}}
  \caption{The Parisi overlap matrix at two temperatures.  The single temperature form of figure \ref{peter-fig1} must be replaced by an $(n+n')\times(n+n')$ matrix which has fluctuating block sizes. In addition it has block diagonal parts that represent overlaps between replicas at the same temperature and off diagonal parts that represent overlaps between the two different temperatures. }
 \label{replic-bis}
\end{figure}

It is difficult to visualise block sizes with real and imaginary parts: in figure \ref{replic-bis} one should imagine that the fluctuations of the block sizes represent the fluctuations of their imaginary part while the real parts of these block sizes can remain fixed.

\section{The case of a general density \texorpdfstring{$\rho(E)$}{} and the replica approach}
 We consider now the case of a general density $\rho(E)$. 
 (For technical reasons, we will assume that $\int \rho(E) dE= \infty$  to ensure  that $Z(\beta)\neq 0$ with probability $1$.  We will also assume that the average partition function $\langle Z(\beta) \rangle$  is finite at least for large enough $\beta$ and of the form $\langle Z(\beta) \rangle = e^{N \Psi(\beta)}$ in contrast to the exponential densities we considered before for which  $\langle Z(\beta) \rangle =\infty$.)

The replica approach consists usually in computing various quantities such as moments of the partition function or overlaps  for an integer number $n$ of replicas and then to try to extend these expressions to  non-integer values of $n$. Here we will see that both for an  integer and a non-integer number $n$ of replicas, one obtains very similar expressions to those obtained in the previous section for the double exponential density.
\subsection{The replica approach for  integer numbers of replicas}

For integer values of the number of replicas one can establish  the following relations (see Appendix B) which are consequences of (\ref{C1}-\ref{C6}) 
\begin{itemize}
\item
For integer $n >0$
\begin{equation}
\langle Z(\beta)^n \rangle  = \sum_{\mu =1}^n {(n-1)! \over (\mu-1)! \, (n-\mu)!}  \ \langle Z(\mu \, \beta) \rangle \ \langle Z(\beta)^{n-\mu} \rangle
\label{E1}
\end{equation}
\item and integer $n \ge  k \ge 1 $
\begin{equation}
\langle Y_k\,  Z(\beta)^n \rangle  = \sum_{\mu =1}^n {(n-k)! \over (\mu-k)! \, (n -\mu)!}  \ \langle Z(\mu \, \beta) \rangle \ \langle Z(\beta)^{n-\mu} \rangle 
\label{E2}
\end{equation}
In (\ref{E2}) as well everywhere else in this paper we use the convention that $(-n)!=\infty$ when $n$ is a positive integer.
\item This approach can be extended to obtain similar, but more complicated expressions for $\langle Y_{k,k'} Z(\beta)^n \rangle$. 
\end{itemize}
One can notice the similarities between 
(\ref{E1},\ref{E2})  and
 (\ref{D6},\ref{D7}). Then as in (\ref{D8},\ref{D9},\ref{D11},\ref{D12})
we see, 
by comparing (\ref{E1}) and (\ref{E2}),
that one can write
\begin{equation}
{\langle Y_k\,  Z(\beta)^n \rangle 
\over \langle   Z(\beta)^n \rangle  }= \sum_{\mu=1}^n  {(n-k)! \ (\mu-1)! \over (n-1)! \, (\mu-k)! }
 \ W(\mu) 
\label{E5}
\end{equation}
where the weights $W(\mu)$ are given by
\begin{equation}
W(\mu)= 
 {(n-1)! \over (\mu-1)! \, (n-\mu)!}  \ {\langle Z(\mu \, \beta) \rangle \ \langle Z(\beta)^{n-\mu} \rangle
 \over \langle Z(\beta)^n \rangle}  
\label{E6}
\end{equation}
A replica interpretation of $W(\mu)$ is that if we choose a replica at random out of the $n$ replicas the probability that it is in a block of size $\mu$ is $W(\mu)$. The combinatorial factor in (\ref{E5}) is then given by the probability that the remaining $k-1$ replicas are also chosen from the same block (see Appendix A). Already, at the level of an integer number $n$ of replicas, we see that, as in (\ref{D8},\ref{D11}), the $\mu's$ can  fluctuate and from the expression for $\langle Y_{k,k'} \, Z(\beta)^n  \rangle$ we have found that they are correlated  as they were in the replica calculation (\ref{D12}) above for the double exponential.

\subsection{When the number $n$ of replicas is not an integer}

We are now going to  write {\it exact} expressions which generalize  (\ref{E1}-\ref{E2}) to non-integer values of $n$.
As some expressions may change depending on the range of $n$, we will consider only the case $n<0$.
As shown in the Appendix B, our derivation is based on  the Cahen-Mellin representation (\ref{F4}) of the exponential.

\begin{itemize}
\item one gets for $n<0$  and $n<\mu < 1$
\begin{equation}
\langle Z(\beta)^n \rangle =  
 \int {dy\over 2\pi} 
 {\Gamma(1- \mu - i y) \ \Gamma(\mu-n + i y) \over  \Gamma(1-n)}  \left\langle Z\Big((\mu  + i y)\beta \Big) \right\rangle  \ \langle Z(\beta)^{n-\mu-i y} \rangle 
\label{E10}
\end{equation}
\item
For $ n < \mu<k$
\begin{equation}
\langle  Y_k \, Z(\beta)^n \rangle =  
 \int {dy\over 2\pi} 
{\Gamma(k- \mu - i y) \ \Gamma(\mu-n + i y) \over  \Gamma(k-n)}  \left\langle Z\Big((\mu  + i y)\beta \Big) \right\rangle  \ \langle Z(\beta)^{n-\mu-i y} \rangle 
\label{E11}
\end{equation}
One can notice the similarity between (\ref{E1},\ref{E2}) and (\ref{E10},\ref{E11}): the ratios of factorials become ratios of Gamma functions and the sums over $ \mu$ become complex integrals over $y$.

For large $N$ (here $N$ is the system size in, for example, (\ref{B13})),  the integrals (\ref{E10}-\ref{E11})
are dominated by a single saddle point at $y=0$ and $\mu=\beta_c/\beta$ when the distribution of energies is of the form (\ref{B13}) and one recovers (\ref{B10},\ref{B11}). One can in principe calculate all the finite size corrections by analysing the neighbourhood of this saddle point.
On the other hand, when the energies take only integer values, there are several saddle points  at the same height
which contribute at $\mu=(\beta_c+ 2\pi i p)/\beta$  
and $\mu'=(\beta_c+ 2\pi i p')/\beta$. By analogy with the case of double exponential of section 4, one can say that the sizes of the blocks in the Parisi matrix fluctuate and take complex values.
In fact it is easy to see  that the prefactors corresponding to these different saddle points in (\ref{E10},\ref{E11}) coincide with the weights $W_p$ in (\ref{D22}).

\end{itemize}

\section{Conclusion}

In this paper we have discussed a number of examples of systems exhibiting one step replica symmetry breaking but for which the original Parisi scheme has to be modified. The simplest case is  a Poisson REM where the density of energies is a sum of two exponentials (see section 4). Our approach is based on developing a recursion relation between the negative moments of the partition function (see  for example  (\ref{D6})). The overlaps can also be expressed in terms of the same negative moments as the partition function (\ref{D7}). Then taking the ratio we obtain expressions (\ref{D8},\ref{D9}) for the overlaps as a weighted sum of two overlap expressions of the form (\ref{B10}) for the original REM. We show that the replica interpretation of this is that the block size of the matrix of overlaps is no longer fixed but each block can choose one of two possible block sizes, $\mu_1$ and $\mu_2$. The Parisi  matrix is still block diagonal, but the block sizes fluctuate, taking either size $\mu_1$ or size $\mu_{2}$ as illustrated in Figure \ref{peter-fig2}. The weights $W_i$  in formula (\ref{D9}) can be interpreted as the probability of a replica chosen at random being in a block of size $\mu_i$. 

This approach is easily extended to other densities of energies that are sums of an arbitrary number of exponentials. This allows us to address REMs that have a disorder distribution that can be represented as a sum of exponentials. An example is the case of finite-size corrections to the REM where we show that the block sizes fluctuate and take complex values. As a consequence, the variance of the block sizes is negative and the corresponding weights are complex. Another example is the REM with discrete energies. Again in the replica approach the block sizes fluctuate and take complex values. We show (see section 4) that this gives a very different form for the distribution $\Pi_\mu(Y)$ (see Figure \ref{peter-fig4}) from the case of Gaussian disorder (see  Figure \ref{figure1}). 

In section 5 we addressed the case of overlaps between two different temperatures. A recursion relation between negative moments of the partition functions allows us to obtain an expression (\ref{F17}) for the overlaps. In the replica approach this can be interpreted as fluctuating block sizes at the two different temperatures but with a relationship (\ref{F8}) between the block sizes. The fluctuations in the block sizes are continuous and imaginary. They are also not unique, with some freedom to chose the real part. However, we expect the overlap expressions and the physical predictions to be the same for each of the choices.  

The REM can be viewed as the paradigmatic model for the one step replica symmetry breaking transition which is known to occur in other spin glass models like the $p$-spin Ising model \cite{Gross_1984_simplest,Gardner_1985_Spin} or the p-spin spherical spin glass \cite{Crisanti_1992_spherical}. The obvious advantage of the REM is that there are exact approaches that do not depend on the replica method. In the original random energy model, with Gaussian disorder, both the replica method and the exact methods produce the same expression (\ref{B10}) for the overlap in the thermodynamic limit. In this case the replica method uses Parisi's replica symmetry breaking scheme with a single step of replica symmetry breaking where the $n$ replicas are partitioned into blocks of fixed size $\mu$. 

Here we saw that rather simple changes in the energy distribution lead to fluctuations of the block sizes of the overlap matrix. It would be interesting to see whether similar effects can be seen in more complex models. We think that the best candidates are systems with discrete energies for which the ground state can be degenerate. These include the directed polymers problem on a tree \cite{Derrida_1988_Polymers} when the the energies on each bond take integer values,
diluted  mean field spin glass models with interactions $J_{ij}= \pm 1$ \cite{Obuchi_2009_Complex}, 
the binary perceptron \cite{Gardner_1988_Optimal,Gardner_1989_Three,Krauth_1989_Storage,Huang_2014_Origin,Ding_2019_Capacity} and the K-sat \cite{Monasson_1996_Entropy,Monasson_1997_Statistical,Mezard_2009_Information}. In all these cases, because the ground state has a non-zero probability of being degenerate, the overlaps $\langle Y_k \rangle \neq 1$ at zero temperature in contrast with what one would expect from (\ref{A7}) in the limit $\mu \to 0$, leading to a $\Pi_\mu(Y)$ consisting of a  sum of delta-peaks at values $Y={1 \over p}$ (where $p$ is the degeneracy of the ground state).  One can then wonder whether  this kind of shape would survive at least at low temperature as it does in our Figure
\ref{peter-fig4} or whether it takes the universal shape of Figure  \ref{figure1} as soon as the temperature is non-zero. 

In all the versions of the REM that we considered here, the matrix of overlaps described a one-step RSB. There is no doubt that one can  generalize what we have seen here, i.e. overlap matrices with fluctuating block sizes,   to systems with several steps of RSB, for example by looking at a generalized random energy model  GREM \cite{Derrida_1985_generalization} with discrete energies. Clearly the statistics predicted by the Ruelle cascade \cite{Ruelle_1987_mathematical} would then be modified. A challenging question   would be to see whether the same phenomenon is present in other  models with full RSB.

\subsection*{G\'{e}rard Toulouse}
\textit{A la fin des années 70 Gérard Toulouse, en comprenant l'importance du problème des verres de spins, a entrainé dans son sillage plusieurs jeunes théoriciens qui sont devenus  des leaders mondialement connus.  A nouveau au milieu des années 80, il a été  l'un des premiers en France à encourager ses plus jeunes collègues  à se lancer dans le domaine des sciences cognitives.
Il a ainsi eu une   influence  majeure sur toute une génération de jeunes physiciens.}
\medskip

(At the end of the 1970s, Gérard Toulouse, recognised the importance of the spin glass problem and inspired the interest of several young theorists, who have since become world-renowned physicists. Again in the mid-1980s, he was one of the first in France to encourage his younger colleagues to enter the field of cognitive science. He has thus had a major influence on an entire generation of young physicists.)

\section{Appendix A: The RSB way of computing overlaps} 

\begin{figure}[h]
 \centerline{\includegraphics[width=6cm]{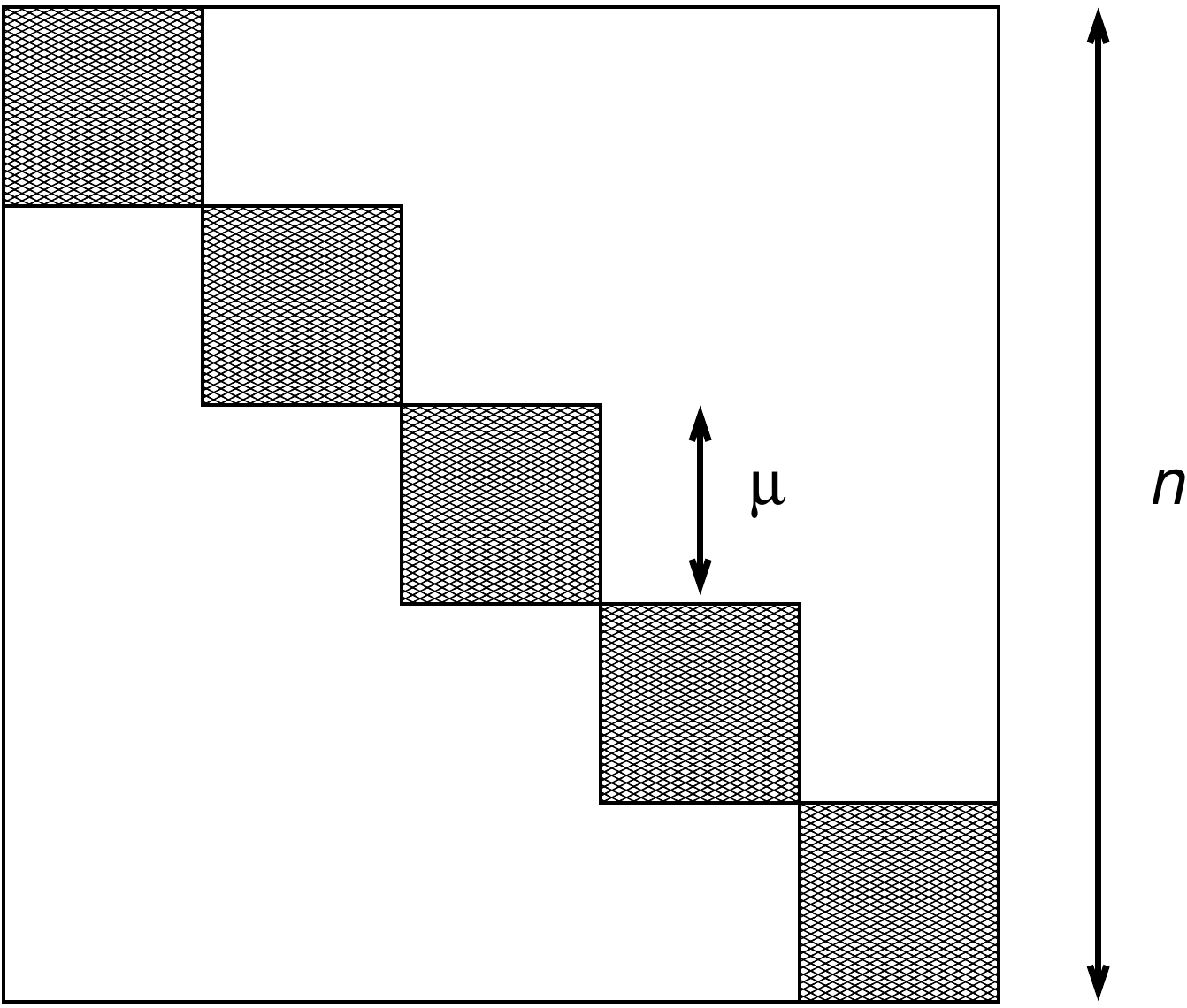}}
 \caption{Parisi's matrix of overlaps between the $n$ replicas: there are ${n \over \mu} $  blocks of $\mu $ replicas. Pairs of replicas inside a block have an overlap larger than $Q$ and pairs of replicas in different blocks have an overlaps  less than $Q$.}
\label{peter-fig1}
 \end{figure}

This appendix gives the  derivation of (\ref{A7},\ref{B10},\ref{B11}) based on replicas.  In the RSB approach \cite{Mezard_1984_Nature,Mezard_1984_Replica,Mezard_1987_Spin}, one considers $n$ configurations called replicas, and given an overlap $Q$, one assumes that these replicas are organized into ${n\over \mu}$ blocks of $\mu$ replicas. All pairs of replicas have an overlap larger than $Q$ if they belong to the same block and an overlap less than $Q$ if they    belong to different blocks. This structure can be represented by the famous $n \times n$ Parisi's matrices  shown in Figure \ref{peter-fig1}.  

According the the  RSB scheme \cite{Mezard_1984_Replica},  the average $\langle Y_k \rangle $   is, in the limit $n \to 0$,  the probability that,  $k$ different replicas  chosen among the $n$  replicas belong all to the same block  

\begin{equation}
\langle  Y_k \rangle = \lim_{n \to 0} \left[ {n \over \mu} \, {\mu(\mu-1) (\mu-2) \cdots (\mu-k+1) \over n (n-1)(n-2) \cdots (n-k+1)} \right]
\label{a1}
\end{equation}
while $\langle Y_{k,k'} \rangle$ is the probability that, choosing $k+k'$ different replicas, all  the first $k$  ones belong to one block and the last $k'$ ones  belong to a different block
\begin{equation}
\langle  Y_{k,k'} \rangle = \lim_{n \to 0} \left[ {n \over \mu} \, \left({n \over \mu}-1 \right) \, 
{\Big( \mu(\mu-1)  \cdots (\mu-k+1) \Big)
\, \Big( \mu(\mu-1)  \cdots (\mu-k'+1) \Big)
 \over n (n-1) \cdots (n-k-k'+1)} \right]
\label{a2}
\end{equation}
and more generally for $p$ groups
\begin{equation}
\langle  Y_{k_1, \cdots k_p} \rangle = \lim_{n \to 0} \left[ (-)^p   {\Gamma\left(p-{n \over \mu} \right) \over \Gamma\left(-{n \over \mu}\right)}\,  {\Gamma(-n) \over \Gamma (k_1 + \cdots k_p-n)} \; \prod_{i=1}^p {\Gamma( k_i-\mu ) \over \Gamma(- \mu)} \right]
\label{a3}
\end{equation}
One can then obtain the expressions (\ref{A6}) using the fact that
\begin{equation}
\langle Y^2 \rangle= \langle Y_4 \rangle + \langle Y_{2,2} \rangle  \ \ \ \ \ ; \ \ \ \ \
\langle Y^3 \rangle= \langle Y_6 \rangle + 3 \langle Y_{4,2} \rangle   + \langle Y_{2,2,2} \rangle
\label{a4}
\end{equation}
(which can be understood  for $\langle Y^2 \rangle$ by noticing that for replicas  1,2 to be  in the same group and replicas 3,4 to be  also in the same group, either 1,2,3,4 are all in the same group or the pair 1,2 and the pair 3,4 belong to different groups. This reasoning can easily be generalized to establish the expression of $\langle Y^3 \rangle$  in (\ref{a4}).)

\section{Appendix B : Derivation of various identities of sections 5 an 6}

\subsection{  An integer number of replicas at a single temperature: derivation of (\ref{E1},\ref{E2})}

One way  to   establish (\ref{E1}) is to expand (as a formal series) in powers of $t$  the following  relation   (see (\ref{C1})) 
$$\phi'(t)  \langle e^{-t Z(\beta)} \rangle =    {d \over dt} \langle e^{-t Z(\beta)} \rangle $$
given that  (see (\ref{C2})) 
$$\phi'(t) = \sum_{\mu=1}^n { (-)^\mu  \ t^{\mu-1} \over (\mu-1)!} \langle Z(\beta \mu) \rangle + O(t^{n})$$
Similarly, using the fact that 
$$\langle Y_k \, Z(\beta)^k e^{-t Z(\beta)} \rangle = \langle Z(k \beta) e^{-t Z(\beta)} \rangle = \Phi_k(t) \langle e^{-t Z(\beta) } \rangle  $$
and expanding   (\ref{C6}) and $ \langle e^{-t Z(\beta) } \rangle  $
   in powers of $t$ one gets for integer $n \ge 0$ :
$$
\langle Y_k \, Z(\beta)^{k+n} \rangle  = \sum_{\mu=0}^n {n! \over \mu! \, (n-\mu)!}  \langle Z\Big((k+\mu)\beta\Big) \rangle \, \langle Z(\beta)^{n-\mu} \rangle $$
which is the same as (\ref{E2}) if one makes the changes of variables $n\to n-k, \mu \to \mu-k$ and  one uses the convention that $(\mu-k)!= \infty$ for $\mu < k$.


\subsection{  A non-integer negative  number of replicas at a single temperature: derivation of (\ref{E10}-\ref{E11})}
One can rewrite (\ref{C3}) as
$$ \langle Z(\beta)^n \rangle = {-1 \over \Gamma(1-n)} \int_0^\infty t^{-n} dt \, \phi'(t) \, \langle e^{-t Z(\beta)} \rangle  \ \ \ \ \  \text{for}  \ \ n<0 
$$
Using first the identity (\ref{F4}) valid for real $\nu>0$
$$ \langle Z(\beta)^n \rangle = {-1 \over \Gamma(1-n)}\int_{-\infty}^\infty  {dy\over2\pi} \int_0^\infty t^{-n} dt \, \phi'(t) \, \Gamma(\nu+iy)t^{-\nu-iy} \langle Z(\beta)^{-\nu-iy}\rangle$$ 
 then an expression of $\phi'(t)$ coming from (\ref{C2}) and then 
 choosing $$0< \nu <1-n $$ 
to ensure the convergence of the integral over $t$, one gets after integrating over $t$ and then over $E$ 
$$ \langle Z(\beta)^n \rangle = 
\int_{-\infty}^\infty  {dy\over2\pi} 
{\Gamma(-n-\nu -iy +1)\, \Gamma(\nu+iy)
 \over \Gamma(1-n)}
  \langle Z( n+\nu +iy)\beta) \rangle \, \langle Z(\beta)^{-\nu-iy}\rangle$$ 
which reduces to (\ref{E10}) in terms of $\mu=n+\nu$.

Similarly from (\ref{C1},\ref{C4}) one has
$$
\langle Y_k Z(\beta)^n \rangle =  {1 \over \Gamma(k-n)} \int_0^\infty t^{k-n-1} dt \, \Phi_k(t)\ \langle  e^{-t Z(\beta)}  \rangle $$
which becomes using
 (\ref{C6}) and (\ref{F4})  for $0<\nu <k-n$
$$ \langle Y_k  Z(\beta)^n \rangle =
\int_{-\infty}^\infty  {dy\over2\pi}
{\Gamma(k-n-\nu -iy +1)\, \Gamma(\nu+iy)
 \over \Gamma(k-n)}
  \langle Z\Big( (n+\nu +iy)\beta\Big)
  \rangle \, \langle Z(\beta)^{-\nu-iy}\rangle$$
and this leads to (\ref{E11})
 in terms of $\mu=n+\nu$.

\bibliographystyle{crunsrt}
\bibliography{Derrida-Mottishaw-30-06-2024}

\def\bysame{\leavevmode ---------\thinspace}
\makeatletter\if@francais\providecommand{\og}{<<~}\providecommand{\fg}{~>>}
\else\gdef\og{``}\gdef\fg{''}\fi\makeatother
\def\cdrandname{\&}
\providecommand\cdrnumero{no.~}
\providecommand{\cdredsname}{eds.}
\providecommand{\cdredname}{ed.}
\providecommand{\cdrchapname}{chap.}
\providecommand{\cdrmastersthesisname}{Memoir}
\providecommand{\cdrphdthesisname}{PhD Thesis}
\begin{thebibliography}{10}

\bibitem{Edwards_1975_Theory}
S.~F. Edwards, P.~W. Anderson, {\og Theory of spin glasses\fg}, \emph{J. Phys. F: Met. Phys.} \textbf{5} (1975), \cdrnumero 5 (en), p.~965.

\bibitem{Parisi_2004_OVERLAP}
G.~Parisi, {\og {THE} {OVERLAP} {IN} {GLASSY} {SYSTEMS}\fg}, in \emph{Stealing the Gold: A celebration of the pioneering physics of {Sam Edwards}} (D.~Sherrington, P.~Goldbart, N.~Goldenfeld, \cdredsname), Oxford University Press, 2004, p.~0.

\bibitem{Sherrington_1975_Solvable}
D.~Sherrington, S.~Kirkpatrick, {\og Solvable model of a spin-glass\fg}, \emph{Phys. Rev. Lett.} \textbf{35} (1975), \cdrnumero 26, p.~1792.

\bibitem{Toulouse_1981_Free}
G.~Toulouse, B.~Derrida, {\og Free energy probability distribution in the {SK} spin glass model\fg}, in \emph{Proceedings of the Sixth Brazillian Symposium on Theoretical Physics} (B.~Ferreira, E.M.;~Koiller, \cdredname), 1981, p.~143-171.

\bibitem{Parisi_1983_Order}
G.~Parisi, {\og Order parameter for spin-glasses\fg}, \emph{Phys. Rev. Lett.} \textbf{50} (1983), p.~1946-1948.

\bibitem{Mezard_1984_Nature}
M.~M{\'e}zard, G.~Parisi, N.~Sourlas, G.~Toulouse, M.~Virasoro, {\og Nature of the spin-glass phase\fg}, \emph{Phys. Rev. Lett.} \textbf{52} (1984), p.~1156-1159.

\bibitem{Mezard_1984_Replica}
M.~M{\'e}zard, G.~Parisi, N.~Sourlas, G.~Toulouse, M.~Virasoro, {\og Replica symmetry breaking and the nature of the spin glass phase\fg}, \emph{J. Phys.} \textbf{45} (1984), \cdrnumero 5, p.~843-854.

\bibitem{Mezard_1987_Spin}
M.~M{\'e}zard, G.~Parisi, M.~A. Virasoro, \emph{Spin glass theory and beyond: {An} Introduction to the Replica Method and Its Applications}, vol.~9, World Scientific Publishing Company, 1987.

\bibitem{Guerra_2003_Broken}
F.~Guerra, {\og Broken replica symmetry bounds in the mean field spin glass model\fg}, \emph{Commun. Math. Phys.} \textbf{233} (2003), \cdrnumero 1, p.~1-12.

\bibitem{Talagrand_2006_Parisi}
M.~Talagrand, {\og The {Parisi} formula\fg}, \emph{Ann. Math.} \textbf{163} (2006), \cdrnumero 1, p.~221-263.

\bibitem{Arguin_2009_structure}
L.-P. Arguin, M.~Aizenman, {\og On the structure of quasi-stationary competing particle systems\fg}, \emph{Ann. Probab.} \textbf{37} (2009), \cdrnumero 3, p.~1080-1113.

\bibitem{Panchenko_2013_Parisi}
D.~Panchenko, {\og The {Parisi} ultrametricity conjecture\fg}, \emph{Ann. Math.} \textbf{177} (2013), \cdrnumero 1, p.~383-393.

\bibitem{Charbonneau_2023_Spin}
P.~Charbonneau, E.~Marinari, M.~Mézard, G.~Parisi, F.~Ricci-Tersenghi, G.~Sicuro, F.~Zamponi (\cdredsname), \emph{Spin glass theory and far beyond}, WORLD SCIENTIFIC, 2023.

\bibitem{Derrida_1987_Statistical}
B.~Derrida, H.~Flyvbjerg, {\og Statistical properties of randomly broken objects and of multivalley structures in disordered systems\fg}, \emph{J. Phys. A: Math. Gen.} \textbf{20} (1987), \cdrnumero 15, p.~5273-5288.

\bibitem{Ghirlanda_1998_General}
S.~Ghirlanda, F.~Guerra, {\og General properties of overlap probability distributions in disordered spin systems. {Towards} {Parisi} ultrametricity\fg}, \emph{J. Phys. A} \textbf{31} (1998), \cdrnumero 46, p.~9149-9155.

\bibitem{Derrida_1985_Sample}
B.~Derrida, G.~Toulouse, {\og Sample to sample fluctuations in the random energy model\fg}, \emph{Journal de Physique Lettres} \textbf{46} (1985), \cdrnumero 6, p.~223-228.

\bibitem{Derrida_1980_Random}
B.~Derrida, {\og Random-energy model: {Limit} of a family of disordered models\fg}, \emph{Phys. Rev. Lett.} \textbf{45} (1980), \cdrnumero 2, p.~79-82.

\bibitem{Derrida_1981_Random}
B.~Derrida, {\og Random-energy model: {An} exactly solvable model of disordered systems\fg}, \emph{Phys. Rev. B} \textbf{24} (1981), \cdrnumero 5, p.~2613-2626.

\bibitem{Eisele_1983_third}
T.~Eisele, {\og On a third-order phase transition\fg}, \emph{Commun. Math. Phys.} \textbf{90} (1983), \cdrnumero 1, p.~125-159.

\bibitem{Galves_1989_Fluctuations}
A.~Galves, S.~Martinez, P.~Picco, {\og Fluctuations in {Derrida}'s random energy and generalized random energy models\fg}, \emph{J. Stat. Phys.} \textbf{54} (1989), \cdrnumero 1, p.~515-529.

\bibitem{Olivieri_1984_existence}
E.~Olivieri, P.~Picco, {\og On the existence of thermodynamics for the random energy model\fg}, \emph{Commun. Math. Phys.} \textbf{96} (1984), \cdrnumero 1, p.~125-144.

\bibitem{Bovier_2002_Fluctuations}
A.~Bovier, I.~Kurkova, M.~L{\"{o}}we, {\og Fluctuations of the free energy in the {REM} and the p-spin {SK} models\fg}, \emph{Ann. Probab.} \textbf{30} (2002), \cdrnumero 2, p.~605-651.

\bibitem{Bolthausen_2007_Random}
E.~Bolthausen, {\og Random media and spin glasses: {An} introduction into some mathematical results and problems\fg} (E.~Bolthausen, A.~Bovier, \cdredsname), Springer, Berlin, Heidelberg, 2007, p.~1-44 (en).

\bibitem{Kistler_2015_Derridas}
N.~Kistler, {\og {Derrida}'s random energy models\fg}, in \emph{Correlated random systems: five different methods}, Springer, 2015, p.~71-120.

\bibitem{Pastur_1989_limit}
L.~A. Pastur, {\og A limit theorem for sums of exponentials\fg}, \emph{Mathematical notes of the Academy of Sciences of the USSR} \textbf{46} (1989), \cdrnumero 3, p.~712-716.

\bibitem{BenArous_2005_Limit}
G.~Ben~Arous, L.~V. Bogachev, S.~A. Molchanov, {\og Limit theorems for sums of random exponentials\fg}, \emph{Probab. Theory Rel.} \textbf{132} (2005), \cdrnumero 4 (en), p.~579-612.

\bibitem{Derrida_2023_Random}
B.~Derrida, P.~Mottishaw, V.~Gayrard, {\og Random energy models: {Broken} replica symmetry and activated dynamics\fg}, \cdrchapname~Chapter 31, p.~657 677, 2023.

\bibitem{Campellone_1995_Some}
M.~Campellone, {\og Some non-perturbative calculations on spin glasses\fg}, \emph{J. Phys. A: Math. Gen.} \textbf{28} (1995), \cdrnumero 8, p.~2149-2158.

\bibitem{Campellone_2009_Replica}
M.~Campellone, G.~Parisi, M.~A. Virasoro, {\og Replica method and finite volume corrections\fg}, \emph{J. Stat. Phys.} \textbf{138} (2009), \cdrnumero 1-3, p.~29-39.

\bibitem{Derrida_2015_Finite}
B.~Derrida, P.~Mottishaw, {\og Finite size corrections in the random energy model and the replica approach\fg}, \emph{J. Stat. Mech: Theory Exp.} \textbf{2015} (2015), \cdrnumero 1, p.~P01021.

\bibitem{Moukarzel_1991_Numerical}
C.~Moukarzel, N.~Parga, {\og Numerical complex zeros of the random energy model\fg}, \emph{Physica A} \textbf{177} (1991), \cdrnumero 1-3, p.~24-30.

\bibitem{Derrida_1991_zeroes}
B.~Derrida, {\og The zeroes of the partition function of the random energy model\fg}, \emph{Physica A} \textbf{177} (1991), \cdrnumero 1-3, p.~31-37.

\bibitem{Saakian_2000_Random}
D.~B. Saakian, {\og Random energy model at complex temperatures\fg}, \emph{Phys. Rev. E} \textbf{61} (2000), \cdrnumero 6, p.~6132-6135.

\bibitem{Bunin_2023_Fisher}
G.~Bunin, L.~Foini, J.~Kurchan, {\og Fisher zeroes and the fluctuations of the spectral form factor of chaotic systems\fg},  (2023), \url{https://arxiv.org/abs/2207.02473}.

\bibitem{Ogure_2005_exact}
K.~Ogure, Y.~Kabashima, {\og An exact analytic continuation to complex replica number in the discrete random energy model of finite system size\fg}, \emph{Prog. Theor. Phys. Supp.} \textbf{157} (2005), p.~103-106.

\bibitem{Jana_2007_Contributions}
N.~Jana, \emph{Contributions to random energy models}, \cdrphdthesisname, Indian Statistical Institute-Kolkata, 2007, \url{https://arxiv.org/abs/0711.1249}.

\bibitem{Derrida_2022_discrete}
B.~Derrida, P.~Mottishaw, {\og The discrete random energy model and one step replica symmetry breaking\fg}, \emph{J. Phys. A: Math. Theor.} \textbf{55} (2022), \cdrnumero 26, p.~265002.

\bibitem{Gardner_1989_probability}
E.~Gardner, B.~Derrida, {\og The probability distribution of the partition function of the random energy model\fg}, \emph{J. Phys. A} \textbf{22} (1989), \cdrnumero 12, p.~1975-1981.

\bibitem{Derrida_1997_random}
B.~Derrida, {\og From random walks to spin glasses\fg}, \emph{Physica D} \textbf{107} (1997), \cdrnumero 2-4, p.~186-198.

\bibitem{Gross_1984_simplest}
D.~J. Gross, M.~M{\'e}zard, {\og The simplest spin glass\fg}, \emph{Nucl. Phys. B} \textbf{240} (1984), \cdrnumero 4, p.~431-452.

\bibitem{Gardner_1985_Spin}
E.~Gardner, {\og Spin glasses with p-spin interactions\fg}, \emph{Nucl. Phys. B} \textbf{257} (1985), p.~747-765.

\bibitem{Parisi_1979_Infinite}
G.~Parisi, {\og Infinite number of order parameters for spin-glasses\fg}, \emph{Phys. Rev. Lett.} \textbf{43} (1979), p.~1754-1756.

\bibitem{Bouchaud_2003_Energy}
J.-P. Bouchaud, F.~Krzakala, O.~C. Martin, {\og Energy exponents and corrections to scaling in {Ising} spin glasses\fg}, \emph{Phys. Rev. B} \textbf{68} (2003), \cdrnumero 22, p.~224404.

\bibitem{Bray_1987_Chaotic}
A.~J. Bray, M.~A. Moore, {\og Chaotic nature of the spin-glass phase\fg}, \emph{Phys. Rev. Lett.} \textbf{58} (1987), \cdrnumero 1, p.~57-60.

\bibitem{Rizzo_2009_Chaos}
T.~Rizzo, {\og Chaos in mean-field spin-glass models\fg}, in \emph{Spin Glasses: Statics and Dynamics} (Basel) (A.~B. de~Monvel, A.~Bovier, \cdredsname), Birkh{\"a}user Basel, 2009, p.~143-157.

\bibitem{Sales_2001_Rejuvenation}
M.~Sales, J.-P. Bouchaud, {\og Rejuvenation in the random energy model\fg}, \emph{EPL} \textbf{56} (2001), \cdrnumero 2 (en), p.~181.

\bibitem{Krzakala_2002_Chaotic}
F.~Krzakala, O.~Martin, {\og Chaotic temperature dependence in a model of spin glasses\fg}, \emph{Eur. Phys. J. B} \textbf{28} (2002), \cdrnumero 2, p.~199-208.

\bibitem{Kurkova_2003_Temperature}
I.~Kurkova, {\og Temperature {Dependence} of the {Gibbs} {State} in the {Random} {Energy} {Model}\fg}, \emph{J. Stat. Phys.} \textbf{111} (2003), \cdrnumero 1 (en), p.~35-56.

\bibitem{Pain_2021_Two}
M.~Pain, O.~Zindy, {\og Two-temperatures overlap distribution for the {2D} discrete {Gaussian} free field\fg}, \emph{Annales de l'Institut Henri Poincaré, Probabilités et Statistiques} \textbf{57} (2021), \cdrnumero 2, p.~685-699.

\bibitem{Derrida_2021_One}
B.~Derrida, P.~Mottishaw, {\og One step replica symmetry breaking and overlaps between two temperatures\fg}, \emph{J. Phys. A: Math. Theor.} \textbf{54} (2021), \cdrnumero 4, p.~045002.

\bibitem{Paris_2001_Asymptotics}
R.~B. Paris, D.~Kaminski, \emph{Asymptotics and {Mellin}-{Barnes} integrals}, Cambridge University Press, 2001.

\bibitem{Crisanti_1992_spherical}
A.~Crisanti, H.-J. Sommers, {\og The spherical p-spin interaction spin glass model: the statics\fg}, \emph{Zeitschrift f{\"u}r Physik B Condensed Matter} \textbf{87} (1992), \cdrnumero 3 (en), p.~341-354.

\bibitem{Derrida_1988_Polymers}
B.~Derrida, H.~Spohn, {\og Polymers on disordered trees, spin glasses, and traveling waves\fg}, \emph{J. Stat. Phys.} \textbf{51} (1988), \cdrnumero 5, p.~817-840.

\bibitem{Obuchi_2009_Complex}
T.~Obuchi, Y.~Kabashima, H.~Nishimori, {\og Complex replica zeros of ±{J} {Ising} spin glass at zero temperature\fg}, \emph{J. Phys. A: Math. Theor.} \textbf{42} (2009), \cdrnumero 7 (en), p.~075004.

\bibitem{Gardner_1988_Optimal}
E.~Gardner, B.~Derrida, {\og Optimal storage properties of neural network models\fg}, \emph{J. Phys. A} \textbf{21} (1988), \cdrnumero 1, p.~271-284.

\bibitem{Gardner_1989_Three}
E.~Gardner, B.~Derrida, {\og Three unfinished works on the optimal storage capacity of networks\fg}, \emph{J. Phys. A} \textbf{22} (1989), \cdrnumero 12, p.~1983-1994.

\bibitem{Krauth_1989_Storage}
W.~Krauth, M.~M{\'e}zard, {\og Storage capacity of memory networks with binary couplings\fg}, \emph{J. Phys. (Paris)} \textbf{50} (1989), \cdrnumero 20, p.~3057-3066.

\bibitem{Huang_2014_Origin}
H.~Huang, Y.~Kabashima, {\og Origin of the computational hardness for learning with binary synapses\fg}, \emph{Phys. Rev. E} \textbf{90} (2014), \cdrnumero 5, p.~052813.

\bibitem{Ding_2019_Capacity}
J.~Ding, N.~Sun, {\og Capacity Lower Bound for the Ising Perceptron\fg}, in \emph{PROCEEDINGS OF THE 51\textsuperscript{ST} ANNUAL ACM SIGACT SYMPOSIUM ON THEORY OF COMPUTING (STOC `19)} (M.~Charikar, E.~Cohen, \cdredsname), 2019, p.~816-827.

\bibitem{Monasson_1996_Entropy}
R.~Monasson, R.~Zecchina, {\og Entropy of the {K}-{Satisfiability} {Problem}\fg}, \emph{Phys. Rev. Lett.} \textbf{76} (1996), \cdrnumero 21, p.~3881-3885.

\bibitem{Monasson_1997_Statistical}
R.~Monasson, R.~Zecchina, {\og Statistical mechanics of the random {K-satisfiability} model\fg}, \emph{Phys. Rev. E} \textbf{56} (1997), \cdrnumero 2, p.~1357-1370.

\bibitem{Mezard_2009_Information}
M.~Mezard, A.~Montanari, \emph{Information, physics, and computation}, Oxford University Press, 2009, 3-22~pages.

\bibitem{Derrida_1985_generalization}
B.~Derrida, {\og A generalization of the random energy model which includes correlations between energies\fg}, \emph{Journal de Physique Lettres} \textbf{46} (1985), \cdrnumero 9, p.~401-407.

\bibitem{Ruelle_1987_mathematical}
D.~Ruelle, {\og A mathematical reformulation of {Derrida}'s {REM} and {GREM}\fg}, \emph{Commun. Math. Phys.} \textbf{108} (1987), \cdrnumero 2, p.~225-239.

\end{thebibliography}

\end{document}